\documentclass[useAMS,usenatbib, usegraphicx]{mn2e}
\usepackage{epsfig}
\usepackage{amsmath}
\usepackage{amssymb}
\usepackage{url}
\usepackage{natbib}
\usepackage{bm}
\newcommand{\kms}{\ensuremath{{\rm km\,s}^{-1}}}
\newcommand{\msun}{\ensuremath{{\rm M}_{\odot}}}

\newcommand{\yr}{\ensuremath{\rm yr}}

\newcommand{\infinity}{\ensuremath{\infty}}

\newcommand{\rmd}{\ensuremath{{\rm d}}}

\newcommand{\msmbh}{\ensuremath{M_{\bullet}}}
\newcommand{\newc}[1]{{#1}}
\newcommand{\new}{}
\title[Finding recoiled BHs in the Milky Way]{Recoiled star clusters
in the Milky Way halo: N-body simulations and a candidate search through SDSS}

\author[O'Leary \& Loeb]{Ryan M.\
  O'Leary$^{1,2}$\thanks{Einstein Fellow, oleary@berkeley.edu} and Abraham
  Loeb$^{2}$\thanks{aloeb@cfa.harvard.edu}\\
$^{1}$Department of Astronomy and Theoretical Astrophysics Center, University of California, Berkeley, CA 94720, USA\\
$^{2}$Department of Astronomy, Harvard University, Cambridge, MA 02138, USA}

\begin{document}
\maketitle

\begin{abstract}
  During the formation of the Milky Way, $\gtrsim 100$ central black
  holes (BHs) may have been ejected from their small host galaxies as
  a result of asymmetric gravitational wave emission. We previously
  showed that many of these BHs are surrounded by a compact cluster of
  stars that remained bound to the BH during the ejection process. In
  this paper, we perform long term $N$-body simulations of these star
  clusters to determine the distribution of stars in these clusters
  today. These numerical simulations, reconciled with our
  Fokker-Planck simulations, show that stellar density profile follows
  a power-law with slope $\approx -2.15$, and show that large angle
  scattering  and tidal disruptions remove $20 - 90\%$ of the stars by
  $\sim 10^{10}\,$yr.  We then analyze the photometric and spectroscopic
  properties of recoiled clusters accounting for the small number of
  stars in the clusters.  We use our results to perform a systematic
  search for candidates in the {\em Sloan Digital Sky Survey}.  We
  find no spectroscopic candidates, in agreement with our expectations
  from the completeness of the survey.  Using generic photometric models
  of present day clusters we identify $\sim 100$
  recoiling cluster candidates. Follow-up spectroscopy would be able to
  determine the nature of these candidates.
\end{abstract}


\section{Introduction}
\label{sec:introduction}

\subsection{Motivation}%%%

In the standard cosmological context of hierarchical galaxy formation,
the first galaxies to form were small ($\sim 10^8\,\msun$) and grew
through subsequent accretion and mergers \citep{loebbook}.  If central
black holes (BHs) were common in the earliest epochs of galaxy
formation, then for most major mergers, the two BHs would also
eventually merge.  If there were any asymmetry in the inspiral and
coalescence of two BHs, whether due to a difference in mass, or the
alignment of the BHs' spin vectors, then there would inevitably be a
net linear momentum kick to the merger remnant
\citep{1962PhRv..128.2471P,1973ApJ...183..657B,1983MNRAS.203.1049F}.
This kick, which is typically hundreds of \kms\ for mergers with
comparable masses
\citep{2006ApJ...653L..93B,2007PhRvL..98w1102C,2007ApJ...659L...5C,2007PhRvD..76f1502T},
is independent of the total mass of the BHs.  Thus, such kicks have
the greatest dynamical effect in the smallest galaxies
\citep{2004ApJ...606L..17M,2006MNRAS.368.1381L,2006MNRAS.372.1540M,2007ApJ...663L...5V,2007ApJ...667L.133S,2008arXiv0805.1420B}. Indeed,
a major merger in the first galaxies would inevitably lead to the
ejection of the BH.  Interestingly, the typical kick velocity is
smaller than the escape velocity of the Milky Way halo, and so,
although the first galaxies to merge would have lost their BHs, many
of these 'rogue' BHs should remain in the Milky Way halo today
\citep{2004ApJ...606L..17M,2005MNRAS.358..913V,2006MNRAS.368.1381L,2008arXiv0805.3154M,SMBHHALO}.

Before coalescence, the BH-BH binary may be surrounded by both a disk
of dense gas and a dense cusp of stars.  For bound matter with orbital
velocities larger than the kick velocity, the gravitational wave kick perturbs the
orbit of the material, but does not unbind it from the recoiled BH
even if the BH is ejected from the galaxy \citep{Loeb}. For gas disks,
viscosity eventually causes the BH to accrete the surrounding gas on
order of a few Myr, leaving the BH as a short lived quasar
\citep{Loeb,2008arXiv0805.1420B,2010arXiv1008.2032G,2010arXiv1008.3313S,2011MNRAS.tmp...38B}.  
After depleting all bound gas, the
BHs will only be visible if they pass through dense gas in the galaxy
and reaccrete material
\citep{2004MNRAS.354..629I,2004MNRAS.354..443I,2005MNRAS.358..913V,2005ApJ...628..873M,2005PhRvD..72j3517B,2008arXiv0805.1420B}.
Stars, on the other hand, are effectively collisionless, and will
remain bound as a long lived system
\citep{2008ApJ...678..780G,2008ApJ...683L..21K,SMBHHALO,2008arXiv0809.5046M},
and may actually be tidally disrupted by the BH after ejection
\citep{2008ApJ...683L..21K} or produce winds 
providing a new source of gas accretion to the black hole \citep{SMBHHALO}.

In O'Leary \& Loeb (2009; hereafter as Paper I), we used $\gtrsim
1000$ merger tree histories of the Milky Way galaxy to calculate the
number and mass distribution of these recoiled BHs. \newc{We assumed
  that after a major merger, with galaxy mass ratio greater than 1:10,
  the black holes merged immediately.}  We found that $\gtrsim 100$
BHs with $\msmbh \gtrsim 10^4\,\msun$ should be in the halo today,
surrounded by compact star clusters that are $\sim 1\%$ of their BH's
mass \citep[see][for a similar discussion applied to the nearby Virgo
  Cluster]{2008arXiv0809.5046M}. \newc{These clusters are expected to
  roughly follow the dark matter distribution of the halo, since they
  have kick velocities typically less than the velocity dispersion of
  the galaxy.  At their typical distances dynamical friction is not important
  over a Hubble time.}  The most massive star clusters have a much
higher \newc{internal} velocity dispersion than globular clusters
because they are gravitationally bound by the BH at their center.
Finding these clusters in the Milky Way halo will allow us to peer
back in time and look at some of the first BHs to form.  In this paper
we re-investigate the long term evolution of recoiled star clusters
using full $N$-body simulations, with a one-to-one correspondence
between stars and $N$-body particles.  We also extend the
Fokker-Planck simulations from Paper I to include large-angle
scattering between stars in order to reconcile these new simulations
with our results from Paper I, and extend the reach of our simulations
to larger BHs.  We use the results of these simulations to generate
the photometric properties of recoiled clusters.  Because these
clusters have so few stars, stochastic variation dominates over the
dispersion of the cluster colors.  Instead of using averaged stellar
evolution tracks of old star systems, we use a Monte-Carlo approach to
generate individual star cluster colors and sizes to identify the
typical properties of these systems and compare them to the properties
of stars and galaxies in the the {\em Sloan Digital Sky
  Survey}\footnote{\url{http://www.sdss.org/dr7/}} Data Release 7
\citep[hereafter SDSS DR7][]{2009ApJS..182..543A}.

Our paper is organized as follows.  In \S~\ref{sec:paperI}, we
describe recoiled clusters and briefly summarize the main results from
Paper I. In \S~\ref{sec:nbody}, we use $N$-body simulations to follow
the long term dynamical evolution of recoiled star clusters.  We then
extend our previous Fokker-Planck analysis in
\S~\ref{sec:fokkerplanck} to include the ejection of stars from large
angle scattering. In \S~\ref{sec:sdss}, we develop a series of simple
photometric models that we use to search for recoiled clusters in
\S~\ref{sec:photosearch}. In \S~\ref{sec:specsearch}, we search the
spectroscopic database of SDSS for candidate clusters.  
Finally in \S~\ref{sec:conclusions}, we summarize our paper and
describe the main conclusions.

\subsection{Stellar Cusps and Recoiled Black Holes}
\label{sec:paperI}
For a relaxed stellar system around a central massive BH,
\citet{1976ApJ...209..214B} found that the stellar density follows a
power-law profile within the radius of influence of the BH, $r_i = G
\msmbh/ \sigma_\star^2$, where $\msmbh$ is the BH mass, and
$\sigma_\star$ is the stellar velocity dispersion.  Out to the radius where
the total mass in stars around the BH is twice the mass of the BH the density profile is
\begin{equation}
  \label{eq:nstar}
  n_\star(r) = \frac{\msmbh}{m_\star}\frac{3-\alpha}{2\pi r_i^3}
  \left(\frac{r}{r_i} \right)^{-\alpha},
\end{equation}
where $\alpha = 1.75$ for a cluster of single mass stars of mass
$m_\star$.  \newc{If the binary black hole merges on a timescale
  comparable to the relaxation time, then the
  \citet{1976ApJ...209..214B} cusp will be regenerated as the binary
  inspirals.} However, if the binary merges on a shorter timescale than
the relaxation time (e.g., through its interaction with gas ), then the stellar
density profile is expected to be much shallower with $\alpha \approx
1$ and with fewer stars within $r_i$ \citep{2006ApJ...648..890M,
  2007ApJ...671...53M}.  For the BH masses we consider here, the
relaxation timescale is much less than the age of the universe,
\begin{equation}
  \label{eq:relax}
  t_r(r) \approx 10^9\,\yr \left(\frac{\msmbh}{10^5\,\msun}\right)^{5/4}\left(\frac{r}{r_i}\right)^{1/4}.
\end{equation}

The kick on the BH remnant occurs over a
timescale much shorter than the orbital time of the stars.  Therefore,
in the frame of the BH, the stars all instantaneously receive a kick,
$-v_{\rm k}$. To first order, stars with a total energy $\lesssim
-m_{\star} v_{\rm k}^2$, will remain bound to the BH as it is ejected
from the galaxy.  For the Keplerian potential of the BH, this
approximately corresponds to all stars with $r \lesssim r_{\rm k} =
(v_{\rm k} / \sigma_{\star})^{-2} r_i$.  From equation~(\ref{eq:nstar}), this
corresponds to a number of stars
\begin{equation}
  \label{eq:nbound}
  N_{\rm cl} \approx \frac{2 \msmbh}{m_\star}
  \left(\frac{v_k}{\sigma_\star}\right)^{2\alpha - 6}
  \\
  \approx 4\times 10^3 \left(\frac{\msmbh}{10^5\,\msun}\right)
  \left(\frac{v_{\rm k}}{5.6\sigma_\star}\right)^{-5/2},
\end{equation}
where we set $\alpha = 1.75$, and used the
$\msmbh-\sigma_\star$ relation to determine $r_i$
\citep{2002ApJ...574..740T}.

The star cluster will begin to expand away from the BH immediately
after it is ejected from the galaxy.  In Paper I, we followed the long
term evolution of the star cluster by numerically integrating the time
dependent, one-dimensional Fokker-Planck equation for stars around a
central BH.  We found that the density cusp of stars around the BH
quickly expands to the point that the relaxation timescale of the
system $t_{\rm r}$ is approximately its age $t_{\rm H}$.  In our
simulations, the total mass in stars was roughly constant. However,
our calculations could not consistently account for either strong
encounters or resonant interactions between stars.  In both cases, we
would expect a larger fraction of stars to be ejected or destroyed.

We estimated the number of recoiled BHs in the Milky Way Halo using an
ensemble of $\gtrsim 10^3$ merger tree histories of the Milky Way,
convolved with the probability distribution of kick velocities
\citep{2007ApJ...662L..63S}. We found that a typical Milky Way like
galaxy contains as many as $100$ recoiled BHs with $\msmbh \gtrsim
10^4\,\msun$, with a mass spectrum $\rmd N / \rmd \msmbh \propto
\msmbh^{-1}$. Because the kick velocity distribution peaks at low
velocities, $\sim 10^2\,\kms$, the majority of recoiled clusters had
the minimal kick velocity needed to escape from the host galaxy. In
Paper I, we estimated this to be $v_{\rm esc} \approx 5.6
\sigma_\star$ immediately after the galaxy merger.  Overall, these results
were consistent with previous studies that looked at a population of
BHs in the Milky Way halo, whether from gravitational wave recoil,
through three body encounters, or as the remnants of the seed
population of black holes \citep{2004ApJ...606L..17M,2005MNRAS.358..913V,2006MNRAS.368.1381L,2008arXiv0805.3154M}

\section{$N$-Body Simulations}
\label{sec:nbody}
Small star clusters around recoiled BHs present an interesting
dynamical system that can be modeled directly in $N$-body simulations,
as well as through approximate methods such as solving the
Fokker-Planck equation.  Because some star clusters have only a few
thousand stars, it is possible to simulate the star clusters with a
one to one correspondence between stars and $N$-body particles.  In
this section, we simulate the star clusters directly, and compare the
results to our previous Fokker-Planck simulations from Paper I.

\subsection{Method}
\label{sec:nbodymethod}
We use the publicly available $N$-body code {\small \sc
  BHint}\footnote{Available at
  \url{http://www.astro.uni-bonn.de/english/downloads.php}}
\citep{bhint} to simulate the long term evolution of star clusters
around recoiled BHs. {\small \sc BHint} was developed to precisely
integrate the equations of motion of stellar systems around massive
BHs, where the BH dominates the motion of the stars.  

{\new The initial conditions for the recoiled BH and star cluster are
  set up by first generating \citet{1976ApJ...209..214B} stellar
  density cusps around a $\msmbh = 10^4\,\msun$ BH, following
  equation~(\ref{eq:nstar}), with $n_\star \propto r^{-1.75}$, such
  that the total mass of stars within $r_i$ is twice that of the black
  hole.  The stars' velocities are initially selected from a Gaussian
  distribution with $\sigma_\star^2 = G \msmbh / r/ (1+\alpha)$, where
  $\alpha = 1.75$, which is an excellent approximation to the velocity
  distribution of stars in a power-law density cusp.  We use two model
  mass functions for the stars. In Model I we use equal-mass stars
  with $m_\star = 1\,\msun$.  In Model II, we use a more realistic
  mass function to model a population of old stars.  Although massive
  stars likely play an important role in the evolution of clusters,
  they are short lived compared to the cluster lifetime.  Stars with
  mass $\lesssim 20\,\msun$, evolve to form neutron stars or white
  dwarfs, which are comparable in mass to long lived main sequence
  stars, and so should not significantly alter the dynamics of the
  system except to increase the mass-to-light ratio\footnote{The total
    amount of mass lost is $<< \msmbh$, and will accordingly only
    perturb the orbits of the stars.}.  More massive stars form black
  holes, with mass $\approx 10\,\msun$, which may dramatically alter
  the dynamics of the system. These black holes may even segregate
  before the binary merges \citep{1993ApJ...408..496M,2009arXiv0909.1318M}.  If
  the black holes dominate the dynamical evolution of the system, such
  systems may not have any luminous stars to observe.  However the
  fraction of black holes in the region immediately after reforming
  the cusp is highly uncertain, and so we take the extreme opposite
  approach and ignore the black hole population. Assuming that there
  are very few black holes, which may be the case in a subset of the
  recoiled clusters, we use a relatively flat mass function for low
  mass stars (${\rm d} N / {\rm d}m_\star \propto m_\star^{-1.35}$)
  with $0.2\,\msun < m_\star < 1.0\,\msun$. The mass of each star is
  generated randomly following this distribution until the total mass
  of the cusp reaches $2\msmbh$.  We use such a shallow power law
  because of the break in the initial mass function at $\approx
  0.5\,\msun$ \citep{2001MNRAS.322..231K}. 

After generating the cusps for Models 1 \& 2, we then kick each star
with a velocity $\bm{v_k} = v_k {\bm{\hat z}}$, and remove all stars
that are unbound to the BH. Approximately $500\,\msun$ of stars remain
bound to the black hole.  With these assumptions we do not account for
stars that are originally unbound to the black hole.  However, unbound
stars do not contribute significantly in numbers deep within a full
cusp.}

We run the simulations for $10^{10}\,$yr, the approximate age of the
clusters.  Stars are removed from the simulation if they are ejected
from the cluster, $E > 0$, if they reach a separation $ a > 10\,$pc,
or if they are tidally disrupted by the central BH\footnote{We have
  also simulated clusters with a much smaller tidal disruption radius
  in order to look at how this may affect the inner density profile.},
$r_{\rm peri} \lesssim r_{\rm tid} = {\rm R}_\star (\msmbh /
m_\star)^{1/3}$.  For each star that is tidally disrupted, we add its
total mass to the BH.  All of our $N$-body simulations follow clusters
with $\msmbh = 10^4\,\msun$ and $v_k = 5.6 \sigma_\star \approx
105\,\kms$. \newc{For all of the reported simulations we set the time
  step criterion $\eta = 0.1$ and the stars orbits were evaluated at
  minimum 80 times per orbit.  We have checked the simulations with
  more precise parameters ($\eta = 0.01$ and 160 evaluations per
  orbit) and found similar results.}  We simulate 40 different cluster realizations for each
Model and average over all the runs.  A typical simulation takes up to
one month on a single core of the Odyssey Cluster at Harvard
University.

\subsection{$N$-Body Results}
\label{sec:nbodyresults}

\subsubsection{Cluster Evolution and Expansion}
\label{sec:clusterevolve}
\begin{figure}
  \centering \includegraphics[width=\columnwidth]{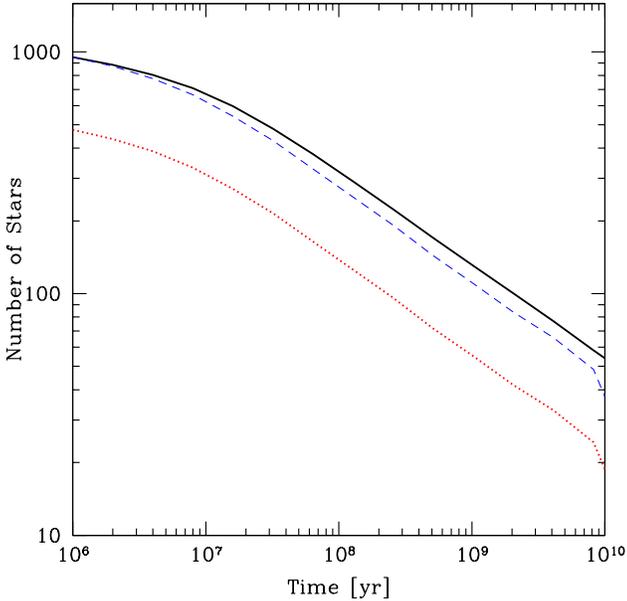}
  \caption{Evaporation of the star cluster in $N$-body
    simulations. The number of stars in the cluster bound to the BH is
    plotted as a function of time.  The solid black line is for the
    cluster with the flat mass function (Model II), the dotted red
    line is the cluster with equal-mass stars (Model I), \newc{and for
    comparison, the dashed blue line is Model I renormalized to have
    the same number of initial stars as Model II.}  After $\sim
    10^7\,$yr the cluster decays as a power-law with time, $N \propto
    t^{-1/2}$.  \newc{Model I has approximately half the number of
    stars of Model II, because the stars are, on average, twice as
    massive.} }
  \label{fig:nstars}
\end{figure}

Immediately after the recoil, the BH is ejected with $\approx~600\,\msun$ of bound stars, in rough agreement with our initial
estimates.  In Figure~\ref{fig:nstars}, we plot the average number of
stars bound to the recoiled BH as a function of time for Models I \&
II. After {\new a} relaxation timescale, $\sim 10^6-10^7\,\yr$, the
star cluster begins to {\new expand 
  and lose stars} as a power-law with $N_{\rm cl}(t) \propto
t^{-1/2}$.  Approximately $40\%$ of the stars are ejected from the
cluster and another $40\%$ of the stars are tidally disrupted by the
BH (see \S~\ref{sec:nbodytidal}). 
\newc{After} both Models begin to evolve, the ratio of the {\em total
  mass} of stars in each Model remains constant in time.

In our simulations there are effectively only 3 parameters that
determine the cluster evolution: the BH mass, $\msmbh$, the average
stellar mass, $m_\star$, and the kick velocity $v_k =
5.6\sigma_\star$.  All other scales in the simulation were determined
through the $\msmbh-\sigma_\star$ relation\footnote{The tidal
  disruption radius depends on the mass ratio of the BH and star, however we found
  that the final number of stars in our simulations was insensitive to the chosen tidal
  disruption radius.}. Since the required kick
velocity to eject the BH scales with $\sigma_\star$, and the total
number of stars in the cluster scales with the BH mass ($\propto
\msmbh$), we can rescale the results of our simulations to find the
number of stars in a recoiled cluster as a function of time ($t_r
\propto \msmbh^{5/4}$) at $t \gtrsim 10^6\,$yr,
\begin{equation}
  \label{eq:nbound-scale}
  N_{\rm cl}(t) \approx 20 \left(\frac{\msun}{<m_\star>}\right)^{3/2}\left(\frac{\msmbh}{10^4\,\msun}\right)^{13/8} \left(\frac{t}{10^{10}\,\yr}\right)^{-1/2},
\end{equation}
for a fixed $v_k = 5.6 \sigma_\star$, and where we have normalized the
scaling relation to match our $N$-body simulations.  The more massive
the BH is, the fewer number of relaxation timescales the cluster will
undergo over a fixed duration in time. Equations~(\ref{eq:nbound}) \&
(\ref{eq:nbound-scale}) are equivalent at $t=10^{10}\,\yr$ when
$\msmbh \approx 2\times 10^6\,\msun$.  For BHs with masses above this
value, the total number of stars in the cluster matches the initial
number of stars at recoil. This has important implications for
recoiled clusters near elliptical galaxies and galaxy clusters
\citep[see][]{2008arXiv0809.5046M}.  For star clusters with $\msmbh
\lesssim 2\times 10^6\,\msun$, such as those we expect around the
Milky Way, the cluster will lose many of its initial stars and have
evolved from its original state. \newc{From
  equation~(\ref{eq:nbound-scale}), and the power-law nature of
  Figure~\ref{fig:nstars}, for a fixed average stellar mass, we expect
  the final number of stars in the cluster will not be significantly
  larger even if the cluster had more stars initially; it would
  instead begin to decay earlier, following the same overall
  functional form $N(t) \propto t^{-1/2}$.}

\begin{figure*}
  \centering \includegraphics[width=\columnwidth]{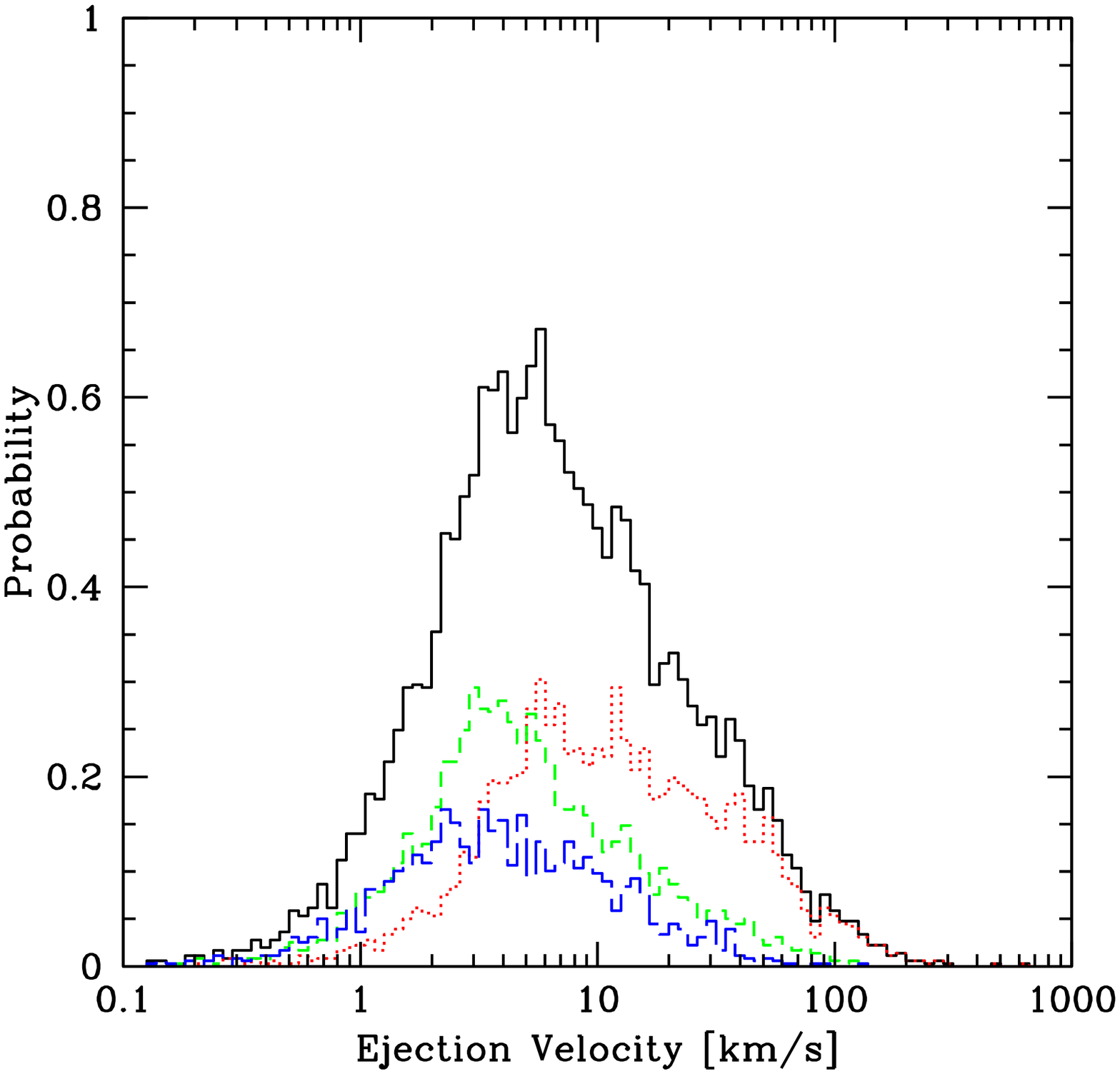}
   \includegraphics[width=\columnwidth]{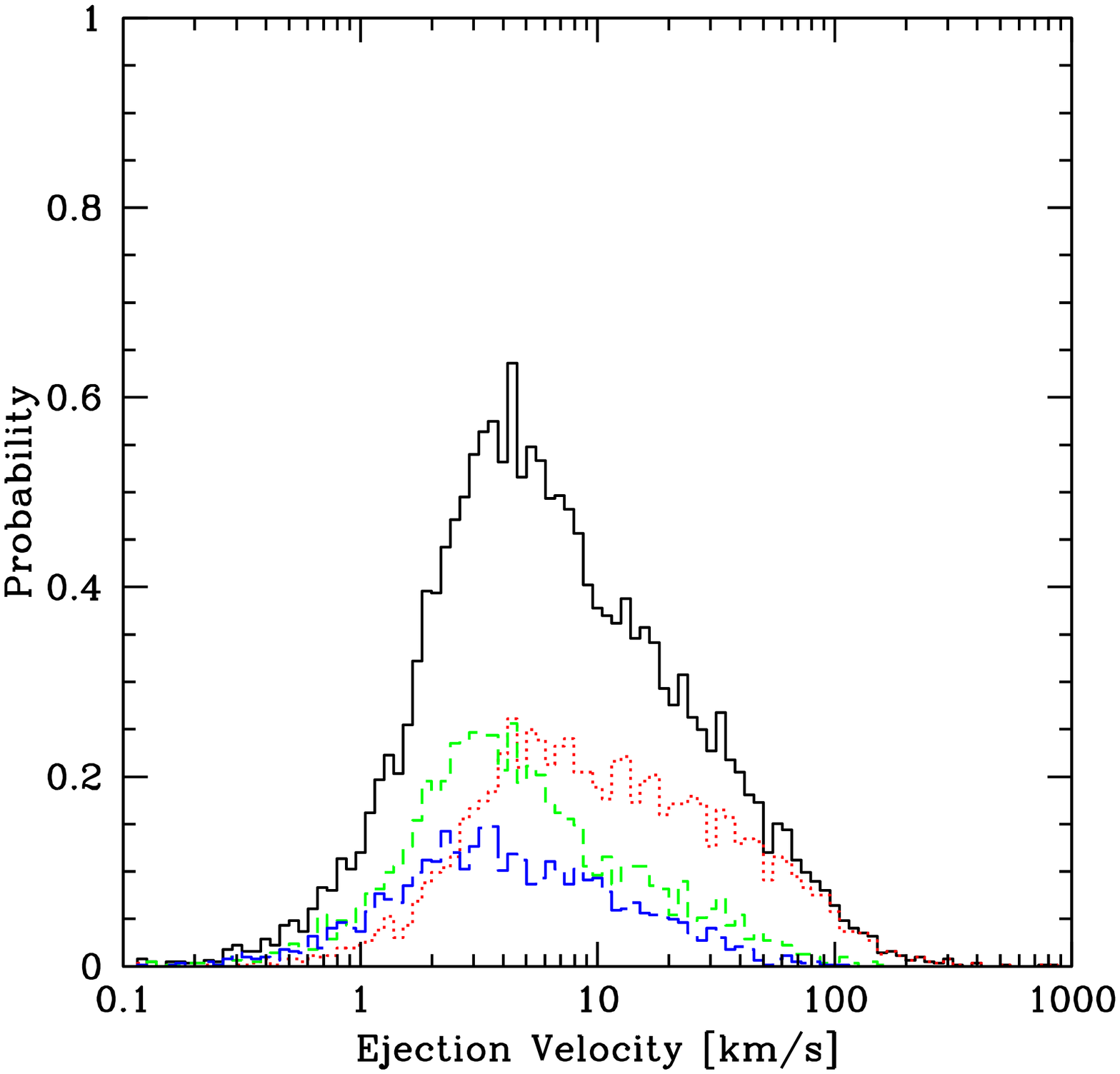}
   \caption{Velocity Spectrum of Ejected Stars.  Plotted is the
     normalized velocity distribution of ejected stars from equal-mass
     clusters (solid black line, left panel) as well as from clusters
     with a flat IMF (solid black line, right panel), Models I and II
     respectively.  The red dotted, green dashed, and blue long-dashed
     lines represent the velocity distribution of the first (ejection
     $\lesssim 10^7\,$yr), second ($\sim 10^7 - 10^8\,$yr), and last
     third ($\gtrsim 10^8\,$yr) of stars \newc{ejected owing to
       two-body scattering}, respectively.  As the cluster expands,
     the typical velocity of ejected stars decreases.  The vast
     majority of stars are ejected with velocities less than the
     velocity dispersion of the Milky Way halo.  The stars with the
     largest ejection speeds are ejected early in the cluster
     evolution, typically at $\lesssim 10^7\,$yr.}
  \label{fig:veject}
\end{figure*}

In Figure~\ref{fig:veject}, we plot the normalized velocity spectrum
of stars ejected from Models I and II as a function of time, looking
at the first, second, and last third of stars ejected from the
cluster. The typical velocity of an ejected star is usually a fraction
of the velocity dispersion of the cluster, which decreases with time
as the cluster expands.  The total velocity distribution of all the
stars ejected from cluster appears almost log-normal,
\newc{independent of the mass function of stars,} with a peak at
$\approx 6\,\kms$, and full-width half maximum an order of magnitude
above the peak value. \newc{On average, nearly twice as many stars
  were ejected from Model II ($\sim 380$ per cluster) compared to
  Model I ($\sim 220$ per cluster). However, the total mass of stars
  ejected is comparable.} From the velocity spectrum, we see that
the slow diffusion of stars to higher energies can not be the cause of
the cluster evaporation. If this were the case, the ejection spectrum
would peak near zero velocity, as most stars become unbound just as
they approach the escape velocity of the cluster.  In fact, the peak
velocity of the ejected stars is near $\Delta v \sim v \sim \sigma$, a
reflection of large angle scattering
\citep{1969A&A.....2..151H,1980ApJ...242..789L}. Indeed, this is
confirmed by the pericenter distance of the stars before they are
ejected, which is always much smaller than the half-mass radius of the
cluster and the boundary of our simulations, $r_{\rm
  peri}<<r_h<<10\,$pc.  Nearly all of the ejected stars will remain
bound to the Milky Way halo ($v < 500\,\kms$). Only a few of the
earliest stars ejected from the cluster have large enough velocities
to constitute a hypervelocity star
\citep{2005ApJ...622L..33B,2003ApJ...599.1129Y}.  This, of course, is
a small fraction of the number of hypervelocity stars that are
expected to be produced during the inspiral of the BH binary
\citep{2003ApJ...599.1129Y}.

{\new After approximately one relaxation timescale, the cluster begins
  to expand as it is heated by the innermost star in the cluster as
  well as by tidal disruptions.  We find in our $N$-body simulations
  that radii that enclose a fixed number of stars scale as a power-law
  $r_N \propto t^{2/3}$.  The same relation is observed in our
  Fokker-Planck simulations (see \S~\ref{sec:fpresults}), in previous
  simulations of black holes in star clusters \citep[see,
    e.g.,][]{2004MNRAS.352..655A}, as well as in previous work
  exploring the expansion of a cluster without a black hole post
  core-collapse \citep[see, e.g.,][ and citing
    articles.]{1961AnAp...24..369H,1984ApJ...280..298G}. The power-law
  index can be obtained by looking at the flow of energy through the
  cluster, so long as the energy is generated in a sufficiently small
  volume.  Following, \cite{2011MNRAS.413.2509G} \citep[see also,][]{1961AnAp...24..369H,1965AnAp...28...62H}, we can analyse the
  flow of energy
\begin{equation}
\label{eq:energyflow}
\frac{\dot{E}}{E} = - 2 \frac{\dot{N}}{N} + \frac{\dot{r}_N}{r_N} = \frac{\zeta}{t_{r}(r_N)},
\end{equation}
at a radius that encloses a fixed number, $N$, of stars. Here,
$\zeta$, is independent of $N$, $r_N$ and $E$.  Under the assumption
that the rate of stars being ejected from the cluster is small
($\dot{N}/N << \dot{r_N}/r_N$), we can solve for $r_N$ as a function
of $t$.  For a cluster in the Keplerian potential of a black hole,
$t_r(r_N) \propto \sigma(r_N)^3 r_N^3 \propto r_N^{3/2}$.  Solving
Eq.~\ref{eq:energyflow}, then yields $r_N \propto t^{2/3}$.  A similar
relation can be found for the expansion of a cluster without a black
hole post core-collapse \citep[see, e.g.,][and citing
  articles.]{1961AnAp...24..369H,1965AnAp...28...62H,1984ApJ...280..298G}, which yields
the same proportionality, however for a different reason ($t_{\rm rh} \propto N^{1/2} r_h^{3/2}$). As the ejection of stars from the cluster becomes important, however, we expect the cluster evolution to deviate from $r_N \propto t^{2/3}$.  }

\subsubsection{Tidal Disruption of Stars \& Resonant Relaxation}
\label{sec:nbodytidal}
\begin{figure}
  \centering \includegraphics[width=\columnwidth]{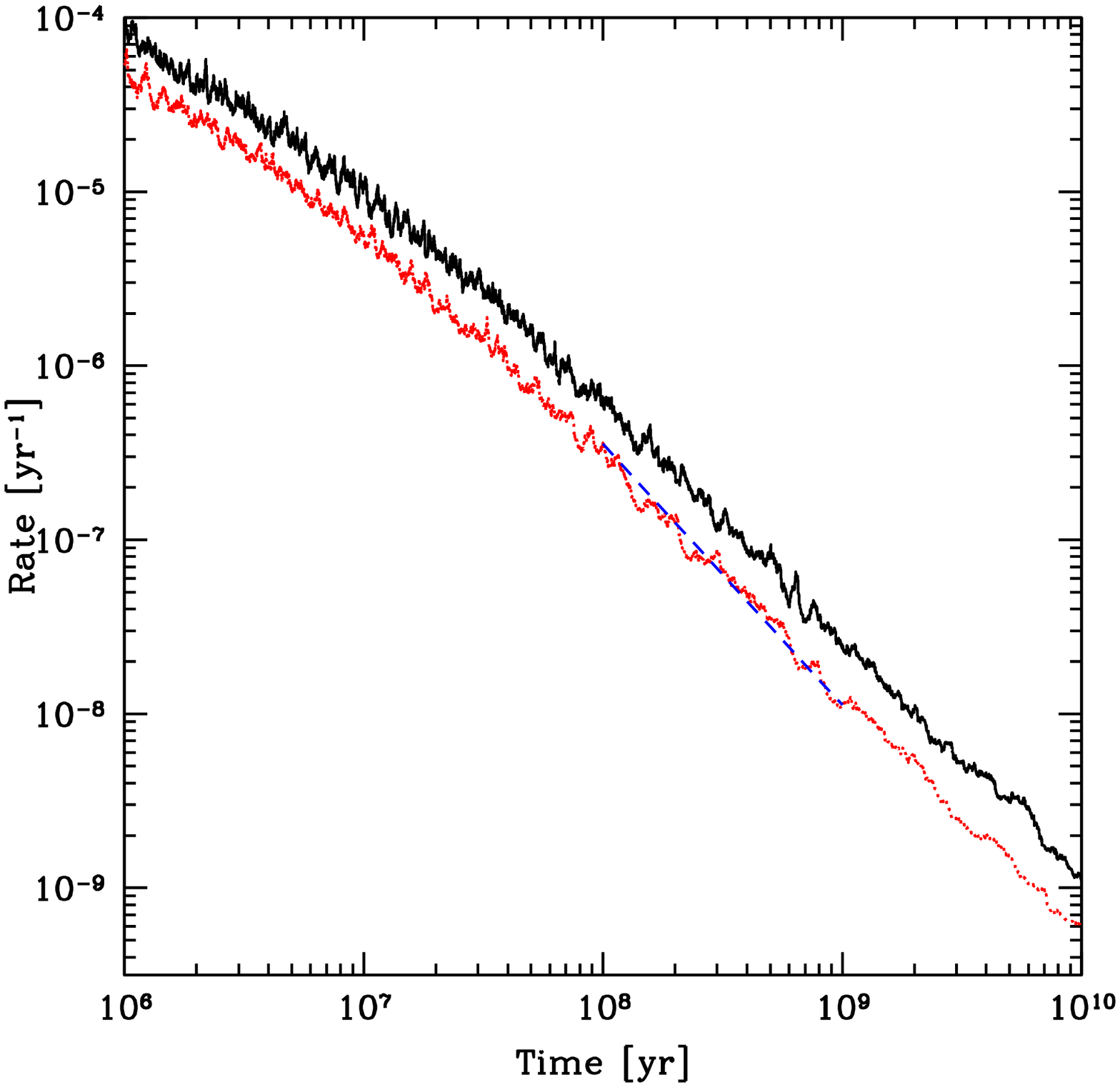}
  \caption{\newc{Tidal disruption rate of stars as a function of time.
      We plot the average tidal disruption rate of the star cluster as
      a function of time since it's central BH was kicked for Models I
      (single mass stars -- dotted red line) \& II (solid black line).
      The long dashed line is a $t^{-3/2}$ power-law drawn for
      comparison.
    For resonant relaxation the tidal disruption rate scales as
    $N/t_{rr}$, which is roughly constant at early times, and then
    scales as $t^{-3/2}$ during the expansion of the cluster.
    Interestingly, although the intrinsic rate of tidal disruption
    events in a given cluster is small, the total rate of all
    recoiling clusters may contribute up to $10^{-6}-10^{-7}\,$yr$^{-1}$, only
    one to two orders of magnitude lower than the disruption rate of
    black holes fixed in galactic nuclei. On the order $\lesssim
    10^4\,\yr$, there is an initial burst of tidal disruptions
    partially because we started with a full loss cone in our
    simulations, however even if the loss cone is empty a burst may
    occur because the kick can put stars into orbits within the loss
    cone \citep{2010arXiv1004.4833S}.}}
  \label{fig:tidal}
\end{figure}

A star will be tidally disrupted by the BH if it comes within a radius
$r_{\rm tid} \sim {\rm R}_\star (\msmbh / M_\star)^{1/3}$. Using this
criterion, we remove stars that are disrupted by the black hole, and
add their mass to the black hole. In Figure~\ref{fig:tidal}, we plot
the average tidal disruption rate of stars in Models I \& II as a
function of time.  {\new  We find that after the first relaxation
  timescale, the time evolution of the disruption rate is well
  approximated as a power-law $\propto t^{-3/2}$.  }

\begin{figure}
  \centering \includegraphics[width=\columnwidth]{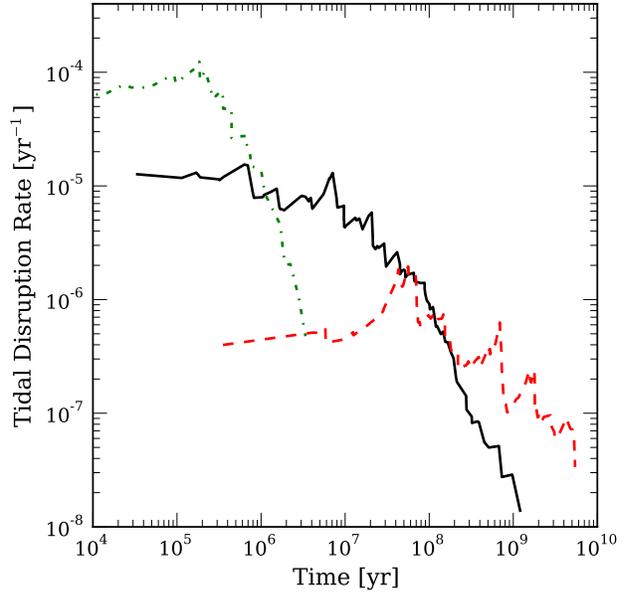}
  \caption{\newc{Tidal disruption rate as a function of stellar mass.  We plot the tidal disruption rate of stellar clusters with $m_\star = 10$, 1, and 0.1$\,\msun$ as green (dash-dotted), black (solid), and red (dashed) lines respectively.  Initially, disruption scales approximately as $\propto m_\star$.  After approximately one relaxation timescale ($\propto m_\star^2$) the cluster begins to expand, and the tidal disruption rate begins to decrease as $\propto t^{-3/2}$.}  There is considerable scatter in these figures because we used only one simulation for each line.}
  \label{fig:tdrr}
\end{figure}

If a star has a small enough angular momentum such that its pericenter
distance is less than the tidal disruption radius, it will be
disrupted in less than one orbital period.  Therefore, the tidal
disruption rate is determined by the { \new flow of stars into the
  empty loss-cone\footnote{The stars can also diffuse to higher
    specific energy, however the fractional change in energy required
    is usually much larger than the fractional change in angular
    momentum.}. The stars can diffuse into the loss-cone 
  through regular two-body relaxation, large-angle scattering,
  or coherent effects such as resonant relaxation
  \citep{1996NewA....1..149R,1998MNRAS.299.1231R}.  We can hope to determine the primary
  mechanism behind loss-cone refilling using the time evolution of the
  system (i.e., $\dot{N}_{td} \propto t^{-3/2} \propto
  N/t$). 

  The diffusion rate of stars into the empty loss-cone from two-body
  relaxation scales approximately as $\sim N/t_r$. After the cluster
  begins to expand, the cluster should expand such that the relaxation
  timescale follows the clusters age,  $t_r \approx t$. Thus the rate
  of tidal disruptions from regular relaxation is $\propto t^{-3/2}$,
  in agreement with our results.  Large-angle scattering will disrupt
  stars with a similar dependence on time, but at a rate reduced by $\sim
  (\ln{\Lambda})^{-1}$.

  For resonant relaxation, the tidal disruption rate should be
  $\approx \gamma N(<r)/t_{rr}(r)$, where $t_{rr}(r)$ is the resonant
  relaxation timescale, and $\gamma$ normalizes the rate and can be
  determined using numerical simulations
  \citep{1998MNRAS.299.1231R,2006ApJ...645.1152H,2008ApJ...683L..21K,2009ApJ...698..641E}.
  If we exclude general relativistic precession, the resonant
  relaxation timescale is determined by the precession of stars from
  their own self-gravity, and scales roughly as $t_{rr} \approx P(r)
  \msmbh/ m_{\star}$, independent of the density profile of the
  stars. For a homologously expanding cluster around a black hole, the
  radius that encloses a fixed number of stars scales as $\propto
  t^{2/3}$ (see \S~\ref{sec:clusterevolve}). Assuming that the orbits
  are nearly Keplerian, $P(r) \propto r^{-3/2}$, the disruption rate
  from resonant relaxation will scale as $N/ (t^{2/3})^{3/2} \propto
  N/t$, the same as for regular relaxation.  As can be seen in
  Figure~\ref{fig:tidal}, the tidal disruption rate scales as $N/t
  \propto t^{-3/2}$ after $t \sim 10^7\,\yr$, when the cluster begins
  to expand.  From scaling arguments alone, we can not determine the
  relative contribution of tidally disrupted stars from resonant
  relaxation or regular relaxation.  Resonant relaxation should be
  more important on these scales since the total enclosed mass in
  stars is much less than the BH mass. However, artificial numerical
  precession can prevent resonant relaxation from occurring in
  numerical simulations.

  To determine the cause of the tidal disruption events we can analyze
  the tidal disruption rates dependence on $m_\star$. For resonant relaxation the disruption
  rate scales as $\propto m_\star$, whereas for regular relaxation the
  rate scales as $t_r \propto m_\star^2$.  Following
  \citet{2008ApJ...683L..21K}, we have performed smaller numerical
  simulations for recoiling star clusters with varying $m_\star$.
  These simulations had only $\approx 200\,\msun$ of initial stars on
  an $n \propto r^{-1}$ density profile.  This profile was chosen so
  that the regular relaxation timescale was shortest at largest radii.
  The tidal disruption rate as a function of time and mass is shown in
  Figure~\ref{fig:tdrr}.  Despite the stochastic variations given the
  small number of stars,  we see that the tidal
  disruption rate scales approximately as $\propto m_\star$ at early
  times, and falls off as $t^{-3/2}$ after approximately one
  relaxation timescale of the system ($\propto m_\star^2$).  We have
  confirmed that the disruptions are indeed caused by resonant effects
  by observing that the stars that are disrupted are preferentially
  from the inner most region of the cusp, and undergo angular momentum
  evolution on a timescale much shorter than the relaxation timescale.

Because the timescale of large-angle scattering ejecting stars is also
proportional to the relaxation timescale, both the tidal disruption
rate and ejection rate have the same functional dependence on time.
By the end of the simulations, approximately $40\%$ of the stars are
disrupted by the BH. This is inconsistent with the results of
\citet{1980ApJ...242..789L} who found that the BHs in globular
clusters are more efficient at ejecting stars from density cusps than
consuming them. However, here we are analyzing only a fraction of the
entire density cusp.}

The tidal disruption of a star from an offset BH is an exciting
possibility for detecting recoiled BHs \citep{2008ApJ...683L..21K}.
Previous work has so far focused on the disruption of stars from
clusters that were recently ejected by merger
\citep{2008ApJ...683L..21K}, as it was thought that the tidal
disruption rate would decline exponentially with time. This is in
contrast with the power-law decline in tidal disruptions found here.
Although the tidal disruption rate peaks early in the cluster
evolution, most recoiled BHs are ejected from their host galaxy in the
early Universe. Taking the time evolution of the clusters from
equation~(\ref{eq:nbound-scale}), the tidal disruption rate of stars in
a cluster today is approximately
\begin{equation}
  \label{eq:trate}
  \dot{N}_{td} \approx 
    10^{-9}\,\yr^{-1} \left(\frac{\msmbh}{10^4\,\msun}\right)^{13/8} \left(\frac{t}{10^{10}\,\yr}\right)^{-3/2}
\end{equation}
for each cluster with $\msmbh\lesssim 2\times10^6\,\msun$ and $t
\gtrsim t_r$.  In Paper I, we found that there are perhaps tens of
clusters with $\msmbh \gtrsim 10^5\,\msun$, and hundreds with $\msmbh
\gtrsim 10^4\,\msun$ around each Milky Way like galaxy.  Thus, per
Milky Way galaxy, the tidal disruption rate of stars by rogue BHs is
approximately $10^{-6}\,\yr^{-1}$ ($10^{-7}\,\yr^{-1}$) for $\msmbh
\gtrsim 10^{5}\,\msun$ ($\gtrsim 10^4\,\msun$).  This is somewhat
lower than the estimated tidal disruption rate of stars by BHs that
reside in galaxies $\sim 10^{-5} - 10^{-6}\,\yr^{-1}$. Upcoming
optical surveys, such as
PTF\footnote{\url{http://www.astro.caltech.edu/ptf/}},
Pan-STARRS\footnote{\url{http://pan-starrs.ifa.hawaii.edu/}}, and
LSST\footnote{\url{http://www.lsst.org/}} are most sensitive to flares
from BHs $\msmbh \sim 10^5 - 10^6\,\msun$ \citep{2009MNRAS.400.2070S},
precisely the range of BHs we expect to have the highest present day
tidal disruption rates. These events can be identified as off-nuclear
flares, which have broad emission lines with a mean redshift
comparable to their nearby galaxies\footnote{Because the recoiling
  clusters are completely in the empty loss cone regime, the typical
  pericenter distance of the tidally disrupted star in a recoiled star
  cluster is always close to $r_{\rm tid}$. This may result in a
  qualitatively different flare than associated with a central black
  hole.}.  Tidal disruption flares from low mass recoiled BHs may be a
promising avenue for detecting these unique systems.

\begin{figure*}
  \centering \includegraphics[width=\columnwidth]{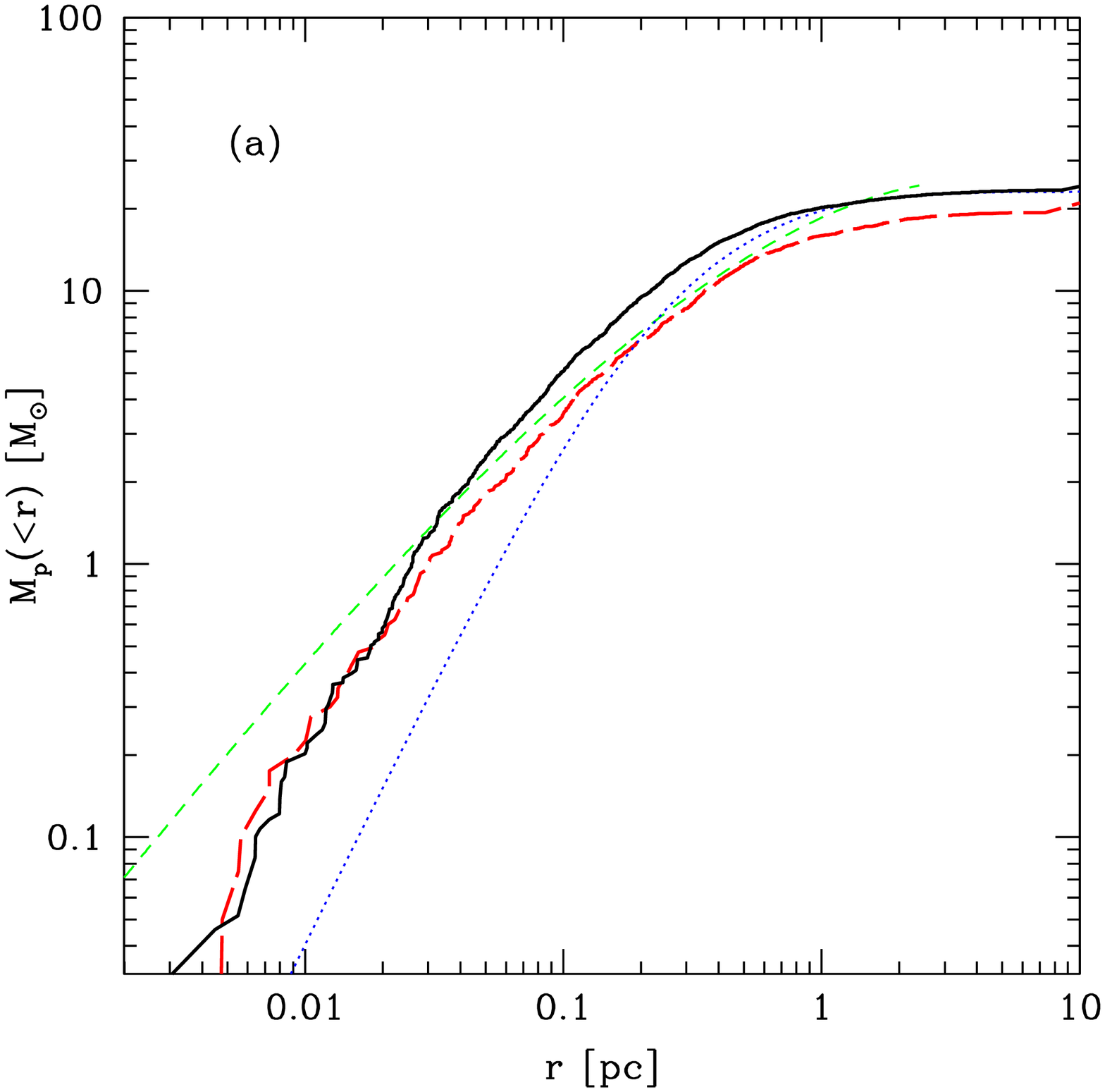}\includegraphics[width=\columnwidth]{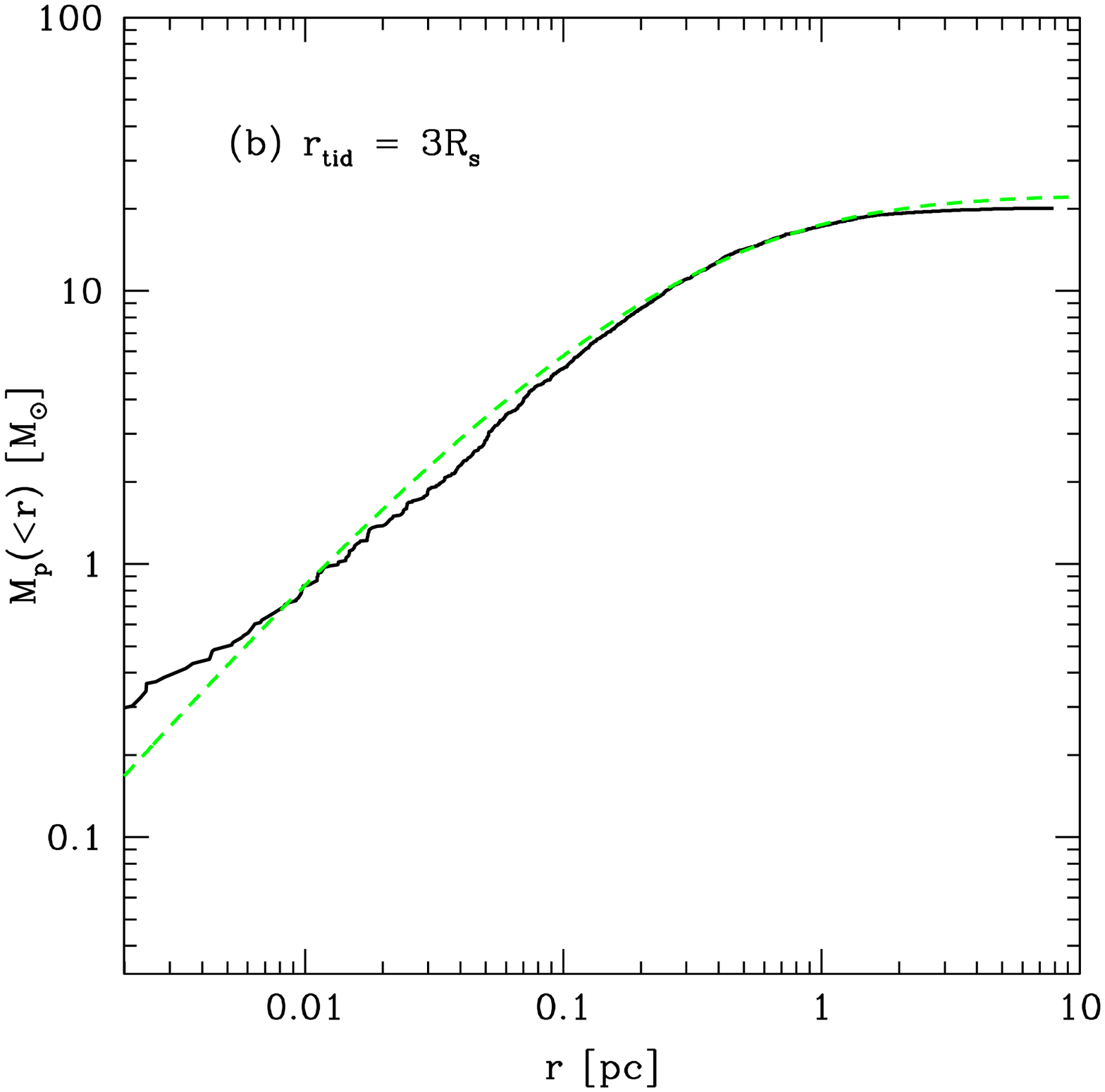}
  \caption{The total projected mass in stars within a distance $r$ of
    the BH. Plotted is the cumulative mass projected within a circle
    of radius $r$ for Model I (single mass: long-dashed red line) \& Model
    II (flat IMF: solid black line) with $\msmbh = 10^4\,\msun$.
    Fig. (a) shows the result of the fiducial $N$-body simulation with
    the appropriate sized tidal disruption radius (see
    \S~\ref{sec:nbodymethod}).   Overlayed, in the green dashed
    line, is the result of our Fokker-Planck simulations from Paper I
    chosen to have the same half-mass radius as our $N$-body
    simulations and  renormalized to have the same total number of stars. \newc{The overall
    shape of the curve matches the $N$-body results remarkably well
    until there is, on average, less than one star enclosed.} \newc{Also
    plotted (blue dotted line) is the result from
    \S~\ref{sec:fokkerplanck} including large angle scattering and
    resonant relaxation.  The line was chosen at the time where the
    total mass of stars in the Fokker-Planck simulation matched the
    $N$-body simulation of Model II.  The Fokker-Planck simulations with resonant
    relaxation tend to clear out too many stars within the half-light
    radius, whereas the simulations without resonant relaxation have
    too many stars and must be rescaled in size in order to match the
    distribution of stars in the simulations. } Fig. (b) shows a similar simulation of Model
    II with a much smaller tidal disruption radius $r_{\rm tid} = 3
    R_s = 6 G \msmbh/c^2$.  The green
    dashed line in Fig. (b) doesn't include two-body scattering. The smaller $r_{\rm tid}$ results in a
    density profile with power-law slope $\alpha = 2.15$ throughout
    the entire cluster, where as, for a more realistic tidal
    disruption radius the density profile flattens within the
    half-mass radius of the cluster.}
  \label{fig:encl}
\end{figure*}

\subsubsection{Present State of Star Clusters}
\label{sec:nbodypresent}

The purpose of these long-term $N$-body simulations is to determine
the present day distribution of stars around recoiled BHs, with the
goal of optimizing the search strategies for these clusters in
\S~\ref{sec:sdss}.  In Figure~\ref{fig:encl}, we plot the average
projected number of stars enclosed within a radius $r$ from the BH. In
addition, we plot the results of our Fokker-Planck simulations from
\S~\ref{sec:fokkerplanck} and Paper I, rescaled in the final number of
stars to match the $N$-body simulations.  The stars in the $N$-body
simulations are distributed with the same functional form and slope
with $\alpha \approx 2.15$ near the half mass radius, as we found in
our initial simulations, despite the total number of stars being
$1/10$ of that found in Paper I.  Regular relaxation, which was
included in Paper I, appears to determine the shape and expansion of
the cluster, whereas a combination of strong-encounters between stars \newc{ and the tidal disruption of stars from resonant relaxation, neither of which were included in our Fokker-Planck simulations of Paper I, determine the final number of stars
in the cluster.}  We conclude from this agreement that recoiled
clusters of comparable mass stars have power-law density profiles with
$\alpha \lesssim 2.15$.

\begin{figure}
  \centering \includegraphics[width=\columnwidth]{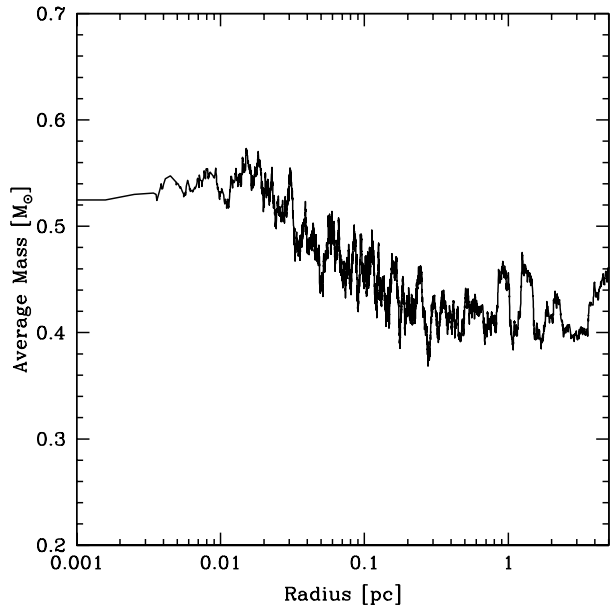}
  \caption{Mass segregation in recoiling clusters. Plotted is the
    average mass of stars as a function of radius when
    $t=6\times10^8\,$yr.  The average mass (with 100 stars per bin)
    slowly declines as a shallow power law out to the half-mass radius
    $r_h \approx 0.2\,$pc.  The lowest mass stars are preferentially
    scattered onto eccentric orbits with larger
    separation.\label{fig:massseg}}
\end{figure}

We find only moderate evidence of mass-segregation in these clusters,
even though the stars spanned a factor of $\sim 10$ in mass, as shown
in Figure~\ref{fig:massseg}.  In our analysis of Model II, we binned
the stars into shells around the BH, and found that the average mass
of stars in each shell decreased as a shallow power-law of radius from
$0.55\,\msun$ and $0.45\,\msun$. Because the massive stars are more
luminous, this segregation will steepen the light density profile in
the cluster.

\newc{To check the robustness of our results on the underlying
  assumptions of our simulations, we have run additional simulations
  that i) start with a shallower density profile, ii) include stellar
  evolution, iii) include general relativistic effects to 2.5
  post-Newtonian order, iv) have a significantly smaller loss-cone,
  or v) use a significantly different mass function. In cases (i) --
  (iv) there was no discernible effect on the stellar density
  distribution or final number of stars in the simulations.  In case
  (v), where the average stellar mass was ten times larger, the
  cluster dissolved in less than $10^{10}\,\yr$. This simulation
  started with 190 stars in orbit around the black hole.  At the end
  of the simulation,  only two bound stars remained.}

\section{Fokker-Planck Simulations}
\label{sec:fokkerplanck}
We found in our direct $N$-body simulations that only $\sim 10\,\%$ of
stars initially bound to a recoiled $10^4\,\msun$ BHs remained after
$10^{10}\,\yr$. In contrast, our time-dependent Fokker-Planck
simulations in Paper I, had little mass loss.  Nevertheless, the
$N$-body simulations and our results from Paper I agree on the shape
and slope of recoiled clusters.  The $N$-body simulations show that
strong encounters between stars as well as an enhanced tidal
disruption rate drive the evaporation of the cluster on timescales
much shorter than standard, uncorrelated perturbative encounters
\citep{1969A&A.....2..151H,1980ApJ...242..789L}.  Here we reintroduce
the time-dependent Fokker-Planck equation for stars around a central
massive object, as originally derived by \citet{1976ApJ...209..214B},
with the addition of \newc{two new sink terms} in order to account for
mass loss caused by strong encounters \newc{and resonant relaxation}.

Following \citep{1976ApJ...209..214B}, we define the relaxation
timescale of the initial cluster to be
\begin{equation}
  \label{eq:relaxtime} t_r = \frac{3(2\pi \sigma_\star^2)^{3/2}}{32 \pi^2 G^2
 m_\star^2 n_\star \ln{\Lambda}},
\end{equation}
where $\sigma_\star$ is the stellar velocity dispersion after the
galaxies merge, $m_\star$ is the average stellar mass, $n_\star$ is
the number density of stars at $r_i = G \msmbh/\sigma_\star^2$, and
$\ln \Lambda \approx \ln (\msmbh/M_\star)$ is the standard Coulomb
logarithm.  In the dimensionless units of time, $\tau = t / t_r$, and
energy, $x = -E/(m_\star \sigma_\star^2)$, the time-dependent
Fokker-Planck equation reduces to {\new
\begin{equation}
  \label{eq:fokkerplanck}
  \frac{\partial g(x,\tau)}{\partial \tau} = -x^{5/2}
  \frac{\partial}{\partial x} Q(x) - R_{\rm lc}(x)-R_{\rm rr}(x) - R_{\rm ss}(x),
\end{equation}}
where $g(x,\tau) = [(2 \pi \sigma_\star^2)^{3/2} n_\star^{-1}]f(E)$
is the dimensionless distribution function of the stars, $Q(x)$ is the
rate at which stars flow to higher energies, $R_{\rm lc}(x)$ is the
tidal disruption rate of stars that diffuse into the BH loss cone \newc{via regular two-body relaxation \citep{1977ApJ...216..883B}},
\newc{ $R_{\rm rr}(x)$ is the tidal disruption rate of stars that fall into the BH loss cone via resonant relaxation \citep{1996NewA....1..149R,1998MNRAS.299.1231R,2006ApJ...645.1152H},} and $R_{\rm
  ss}(x)$ the rate that stars are ejected from the cluster owing to
strong encounters. \citet{1976ApJ...209..214B,1977ApJ...216..883B}
found that
\begin{eqnarray}
  \label{eq:flowrate}
  Q(x) = \int_{-\infinity}^{x_{\rm td}}
  \rmd y [{\rm max} (x,y)]^{-3/2} 
  \left(g(x) \frac{\partial
      g(y)}{\partial y} - g(y)\frac{\partial
      g(x)}{\partial x}\right).
\end{eqnarray}
and
\begin{equation}
  \label{eq:losscone}
  R_{\rm lc}(x) \approx \frac{g(x)^2}{ \ln[J_c(x)/J_{\rm LC}]},
\end{equation}
where $J_c(x)/J_{\rm LC} \approx (x/x_{\rm td})^{1/2}$, and where $x_{\rm td}
\approx (\msmbh / m_\star)^{-1/3} r_i/R_\star$ is the maximum specific
energy of a star before tidal disruption.

In systems with nearly Keplerian orbits, torques between the stars add
coherently, and can efficiently randomize the angular momentum of the
stars on a timescale much shorter than the regular (non-coherent)
relaxation time.  This process, known as resonant relaxation
\citep{1996NewA....1..149R}, can lead to an enhanced rate of tidal
disruptions as the stars enter the loss cone
\citep{1998MNRAS.299.1231R}. We follow \citet{2006ApJ...645.1152H}, who
derived an approximate expression for the resonant relaxation driven
tidal disruption rate for Eq.~\ref{eq:fokkerplanck}.  They found that the rate is approximately 
\begin{equation}
\label{eq:rrloss}
R_{\rm rr}(x) \approx \chi \frac{g(x)}{\tau_{\rm rr}(x)}, 
\end{equation}
where $\chi$ is an unknown efficiency factor of order unity, and
$\tau_{\rm rr}(x)$ is the resonant relaxation timescale.  In our
numerical simulations we do not include general relativistic
precession\footnote{In \S~\ref{sec:nbodypresent}, we have included general relativistic precession and found no quantitative different in the tidal disruption rate or the distribution of stars in the cluster}, so the only form of precession that limits the resonant
relaxation is caused by the enclosed mass of the system. In these
circumstances \citet{2006ApJ...645.1152H} found $\tau_{\rm rr} \approx
.0278 x^{3/2}$.

Although strong encounters are less important in calculating the flow
of stars to higher and lower energies, in eq.~(\ref{eq:flowrate}), a
single strong encounter can eject a star from the cluster.
\citet{1969A&A.....2..151H} first calculated the escape rate of stars
from isolated star clusters for an arbitrary distribution of stars.
\citet{1980ApJ...242..789L} extended this work to stars around a
central point mass. They found that strong encounters are important in
calculating the flux of stars out of the cusp, as confirmed in our
$N$-body experiments in \S~\ref{sec:nbody}. By changing the limits of
integration in equation~(35) of \citet{1980ApJ...242..789L}, we
derive the rate that equal-mass stars are ejected from the cluster as
a function of energy,
\begin{equation}
  \label{eq:ssrate}
  R_{\rm ss}(x) = \frac{3}{2} x^{5/2} \frac{g(x)}{(x-x_0)^2}
  \int \frac{g(y){\rm d}y}{(y+x-x_0)^{3/2}}\frac{1}{\ln \Lambda},
\end{equation}
where $x_0 \lesssim 0 $ is the negative specific energy required to be
ejected. Note that in our dimensionless units,
equation~(\ref{eq:ssrate}) is suppressed by the Coulomb logarithm $(\ln
\Lambda)^{-1}$, compared to the rest of
equation~(\ref{eq:fokkerplanck}).

We determine the time evolution of the cluster by numerically solving
equation~(\ref{eq:fokkerplanck}) with the boundary conditions $g(x<0)
= \exp(x)$ and $g(x>x_{td}) = 0$ until $\tau = 2$, at which point we
set $g(x<0) = 0$. We remove stars from the kick by scaling the
distribution function of stars as $g(x) \rightarrow g(x)
z^{2.5}/(1+z^{2.5})$, where $z = 2 x / (v_k/\sigma_\star)^2$. This
yields an asymptotic density profile with $n \propto r^{-4}$ for $r
\gtrsim r_k$, as expected immediately after the kick
\citep{2008ApJ...683L..21K}.  We use a variety of $x_0 = 0.01$, 0.1,
0.25, 0.5, 1.0, 2.0, 10.0, to explore the importance of $x_0$ in
matching the number of stars in \S~\ref{sec:nbody},\newc{ as well as $\chi =
0.1, 0.5, 0.7, 0.8, 0.9, 1.0, 2.0$ to explore the uncertainty in the
efficiency of resonant relaxation}. In all of our calculations we set
$\ln \Lambda = 10$.

\subsection{Fokker-Planck Results}
\label{sec:fpresults}

\newc{In our calculations, we find reasonable agreement between the
  $N$-body simulations and the numerical solution of the Fokker-Planck
  equation {\emph {only}} when we include a new sink term, $R_{rr}$,
  which accounts for the tidal disruption of stars owing to resonant
  relaxation.  We find the best agreement when we set the resonant
  relaxation parameter $\chi = 0.8$. When we exclude resonant
  relaxation, as we did in Paper I, we only get the proper functional
  form of the $N$-body solution to the density profile, but not the
  proper number of stars.}
\newc{Without resonant relaxation, the Fokker-Planck simulation expand
  too rapidly at the half-mass radius.}  Indeed, when comparing the
radii that enclose the inner 1, 10, or 100 stars of the Fokker-Planck
simulations to the $N$-body simulations we find complete agreement
that the cluster expands self-similarly as $r \propto t^{2/3}$ (see
the discussion in \S~\ref{sec:clusterevolve}).  At the radius that
encloses half of the cluster mass, we find the Fokker-Planck
simulations still scale as $t^{2/3}$, and the $N$-body simulations
scales as $t^{1/3}$.  The $t^{1/3}$ scaling is in agreement with our
expectation that the relaxation timescale should be approximately the
age of the cluster (Paper I).  

{\new Resonant relaxation reconciles the two fundamental differences
  between our previous Fokker-Planck simulations and our current
  $N$-body simulations.  Resonant relaxation destroys stars at a
  faster rate than regular relaxation, reducing the number of stars in
  the cluster. As a consequence of the depleted stars, the outer
  cluster expands more slowly, matching the N-body simulations.  In
  Figure~\ref{fig:encl}, we plot the enclosed mass profile of the
  clusters from our $N$-body simulations, along with our Fokker-Planck
  simulations including and excluding resonant relaxation (with $\chi
  = 0.8$). Resonant relaxation reduces the number of stars in the
  simulation to the correct value. Although including resonant
  relaxation in the Fokker-Planck simulations does give a similar mass
  density profile, there are too few stars in the inner most region.
  Unfortunately, the Fokker-Planck simulations are one-dimensional,
  and can not take into account the anisotropy that must develop as
  the cluster expands, or the preference to deplete eccentric orbits
  near the black hole.}

\begin{figure}
  \centering \includegraphics[width=\columnwidth]{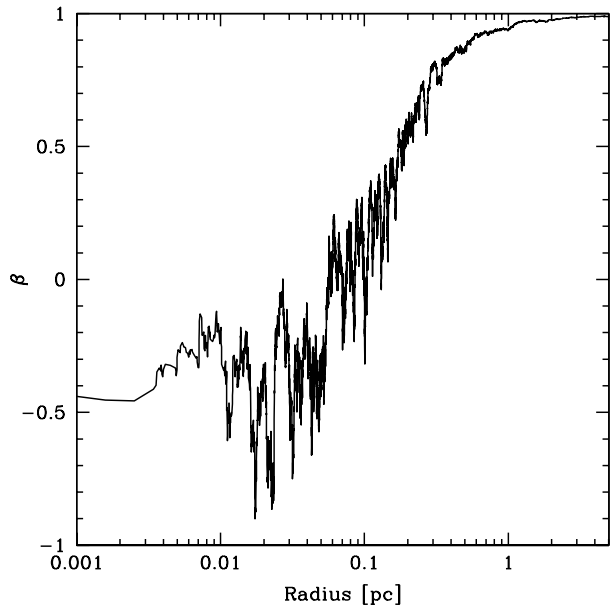}
  \caption{Radial anisotropy of the expanding cluster.  Plotted is the
    anisotropy parameter, $\beta \equiv 1 - \sigma_t^2/\sigma_r^2$ as
    a function of radius \newc{for Model II when} $t=6\times 10^8\,$yr. The
    cluster shows a large degree of anisotropy at nearly all radii.
    From the innermost stars outward, the stars in the cluster move
    from preferentially tangential orbits ($\beta <0$) to radial
    orbits ($\beta \approx 1$). In our Fokker-Planck simulations we
    assume isotropy, i.e., $\beta = 0$, at all radii at all
    times.\label{fig:aniso}}
\end{figure}

\begin{figure}
  \centering \includegraphics[width=\columnwidth]{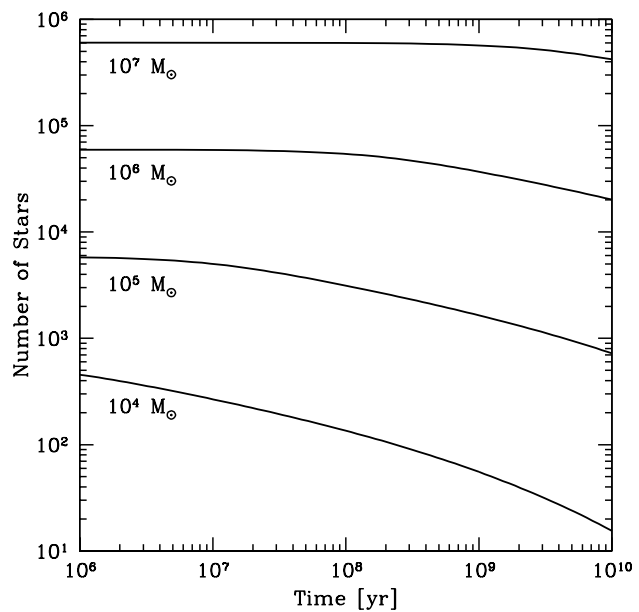}
  \caption{The fate of stars surrounding recoiled BHs. The number of
    stars in the cluster bound to the BH is plotted as a function of
    time for BHs with $\msmbh = 10^7\,\msun$ to $10^4\,\msun$ from top
    to bottom.  \newc{The evaporation of the clusters was calculated with
    the Fokker-Planck simulations including the tidal disruption of
    stars from resonant and regular relaxation ($\chi = 0.8$), as well as the loss
    of stars from strong encounters ($x_0 = 0.1$).  For $\msmbh \gtrsim
    10^6\,\msun$, the cluster loses no more than $\approx 60\,\%$ of
    its original mass.}}
  \label{fig:nvstfpm}
\end{figure}

The anisotropy of the cluster is often measured by the parameter 
\begin{equation}
\beta \equiv 1 - \frac{\sigma_t^2}{\sigma_r^2}
\end{equation}
where $\sigma_r$ is the radial velocity dispersion of the system, and
$\sigma_t$ is the one-dimensional tangential velocity dispersion, such
that $\sigma_\star^2 = \sigma_r^2 + 2\sigma_t^2$.  For an isotropic
cluster, $\sigma_r = \sigma_t$ and $\beta = 0$.  A cluster with stars
on only radial orbits will have $\beta = 1$.  We plot the anisotropy
of the $N$-body simulations as a function of radius in
Figure~\ref{fig:aniso}. The cluster shows a large degree of anisotropy
at nearly all radii.  From the innermost stars outward, the stars in
the cluster change from preferentially tangential orbits ($\beta <0$)
to radial orbits ($\beta \approx 1$).  Indeed, the development of
anisotropy is a natural consequence of the conservation of angular
momentum in a Keplerian potential.  As the cluster expands, the
outermost stars must be on more radial orbits.  The innermost stars,
on the other hand, have their eccentric orbits depleted when they are
tidally disrupted by the BH.  In the our one dimensional Fokker-Planck
simulations, we assume isotropy, which effectively relaxes the angular
momentum of the stellar population on a timescale much shorter than
the actual relaxation timescale. This causes the cluster to disrupt
stars very close to the black hole, that would otherwise remain on
circular orbits.  This scenario can be tested with an appropriate two
dimensional Fokker-Planck code
\citep[e.g.,][]{1980ApJ...242..765C,1995PASJ...47..561T,1999ApJ...518..233D},
extended to include the effects of resonant relaxation.
Despite the anisotropy of the cluster, the density profile predicted
by the Fokker-Planck simulations matches the $N$-body simulations
remarkably well, \newc{except in the innermost region of the cusp}.

The $N$-body simulations of \S~\ref{sec:nbody} are computationally
challenging given the long timescale of the calculation and can not be
easily extended to larger star clusters.  Solving the Fokker-Planck
equations, however, does not depend on $N_{\rm cl}$. Rather, it is
calculated on a fixed grid in energy space, and takes only a short
computational time to complete.  We therefore use the Fokker-Planck
code to compute the evolution of more massive recoiled star clusters
around $\msmbh = 10^5$, $10^6$, and $10^7\,\msun$, all with kick
velocity scaled to their respective velocity dispersion, $v_k =
5.6\sigma_\star$, and normalized by the $\msmbh-\sigma_\star$
relation. \newc{We set the free parameters $\chi = 0.8$ and $x_0 = 0.1$,
  which give comparable time evolution to the $N$-body simulations in
  \S~\ref{sec:nbody}.} In Figure~\ref{fig:nvstfpm}, we plot the number
of stars as a function of time for these recoiled clusters.  We find
that the evolution of star clusters around black holes with $\msmbh
\gtrsim 10^6\,\msun$, lose less than $60\,\%$ of their mass over
$10^{10}\,\yr$.  For the largest black holes with $\msmbh \gtrsim
10^7\,\msun$, the clusters lose negligible mass over the age of the
universe.  We can therefore expect that the most massive clusters
represent the conditions of the stellar cusp when they were recoiled
from their parent galaxy.

We find that the total number of stars at the end of the simulations
is only weakly dependent on the value we choose for $x_0$.  Indeed,
over two orders of magnitude in $x_0$, the final number of stars in
the cluster changed by only $10-20\%$.  This is because for most of
the range of $x_0$ values, the corresponding velocity for stars
ejected with energy $x_0$, was less than mean ejection velocity as
seen in Figure~\ref{fig:veject}.  In the limit $x_0 \rightarrow
\infinity$, equation~(\ref{eq:ssrate}) goes to zero, and we recover
the results of our simulations from Paper I. Unfortunately, we can not
use these simulations to calibrate $x_0$ for BHs in galactic nuclei,
where there is a reservoir of stars outside of $r_i$.  When $x_0
\approx 0$, equation~(\ref{eq:ssrate}) is not accurate because it does
not account for the flux of stars to lower energy states from outside
of the cusp or the return of stars that are still bound to the cusp of
stars \citep{1980ApJ...242..789L}.

\newc{The evolution of the star clusters is sensitive to the value of
  $\chi$.  For sufficiently small, $\chi \lesssim 0.1$, clusters of
  stars around $10^4\,\msun$ BHs only lose $\approx 60\%$ of their
  mass over $10^{10}\,\yr$.  Likewise, for  $\chi \gtrsim 5$,
  the cluster loses mass so rapidly, that it has $\lesssim 1$ star
  after $4\times 10^9\,\yr$.  Such a scenario may represent a cluster
  with high concentration of stellar mass black holes. }

\section{Photometric Properties of Old Clusters}
\label{sec:sdss}
SDSS DR7 \citep{2009ApJS..182..543A} has $\sim 3.6\times 10^8$
unique photometric objects that we would like to sort through in order
to find  $\sim 100$ candidate recoiled star
clusters.
Our goal is to develop the most general photometric model possible for
recoiled star clusters in SDSS, while eliminating false positives as
efficiently as possible.  Generally, we expect the clusters to contain
as few as $\sim 20$ stars for the smallest BHs $\msmbh = 10^4\,\msun$,
up to $\sim 10^4$ stars for the most massive BH in the halo $\msmbh =
5\times 10^5\,\msun$.  These clusters should have a power-law density
profile with $ \alpha \approx 2.15$, but certainly $1.75< \alpha < 2.25$,
which corresponds to a cusp of stars which flows {\em away} from the
BH.  Since the Milky Way has not had a recent major merger, we expect
the clusters to be old.  In this section, we develop such a model,
focusing on the photometric properties of a stochastic cluster of old
stars with a power-law density profile.  Their spectra should indicate
a large $\sigma_\star$ for clusters around the most massive BHs,  and large mass-to-light ratio at redshift
$z=0$, but we focus on photometric identification of candidates for
spectroscopic follow-up.

\subsection{Cluster Models}
\label{sec:models}
We generate model star clusters by randomly selecting stars from a
Kroupa initial mass function of stars \citep{2003ApJ...598.1076K}
using two main modes of star formation. For Model A, we assume that
the stars formed continuously in time with a constant star formation
rate until the BH was ejected after \newc{$\sim 5\times
10^9\,\yr$}. This is consistent with the estimated star formation
history of the MW galactic center
\citep{1999ApJ...520..137A,2003ApJ...594..812G}.  We contrast this
with Model B, where the stars formed simultaneously with the BH
merger, as galaxy mergers are often associated starbursts.  In all
instances we assume that the time between the merger of the galaxy and
the merger of the BHs is negligible.

The precise photometric properties of the stars depends on their
metallicity and age. We consider three different
metallicity histories of stars: (I) solar metallicity ($Z = 0.02$),
(II) sub-solar metallicity ($Z = 2\times 10^{-4}$), and (III) the
estimated time evolution of metallicity of the galactic center.

In all of our calculations we use the online single star population
synthesis code {\small \sc BaSTI}\footnote{Available at
  \url{http://astro.ensc-rennes.fr/basti/synth_pop/index.html}}
\citep{2004ApJ...612..168P,2007AJ....133..468C}, and convert all
observables to the SDSS color system using Table 1 of
\citet{2005AJ....130..873J}. The typical uncertainty in the colors
from our conversion scheme is much smaller than the intrinsic
variation in the colors of the star clusters.

\begin{figure*}
  \centering \includegraphics[width=\columnwidth]{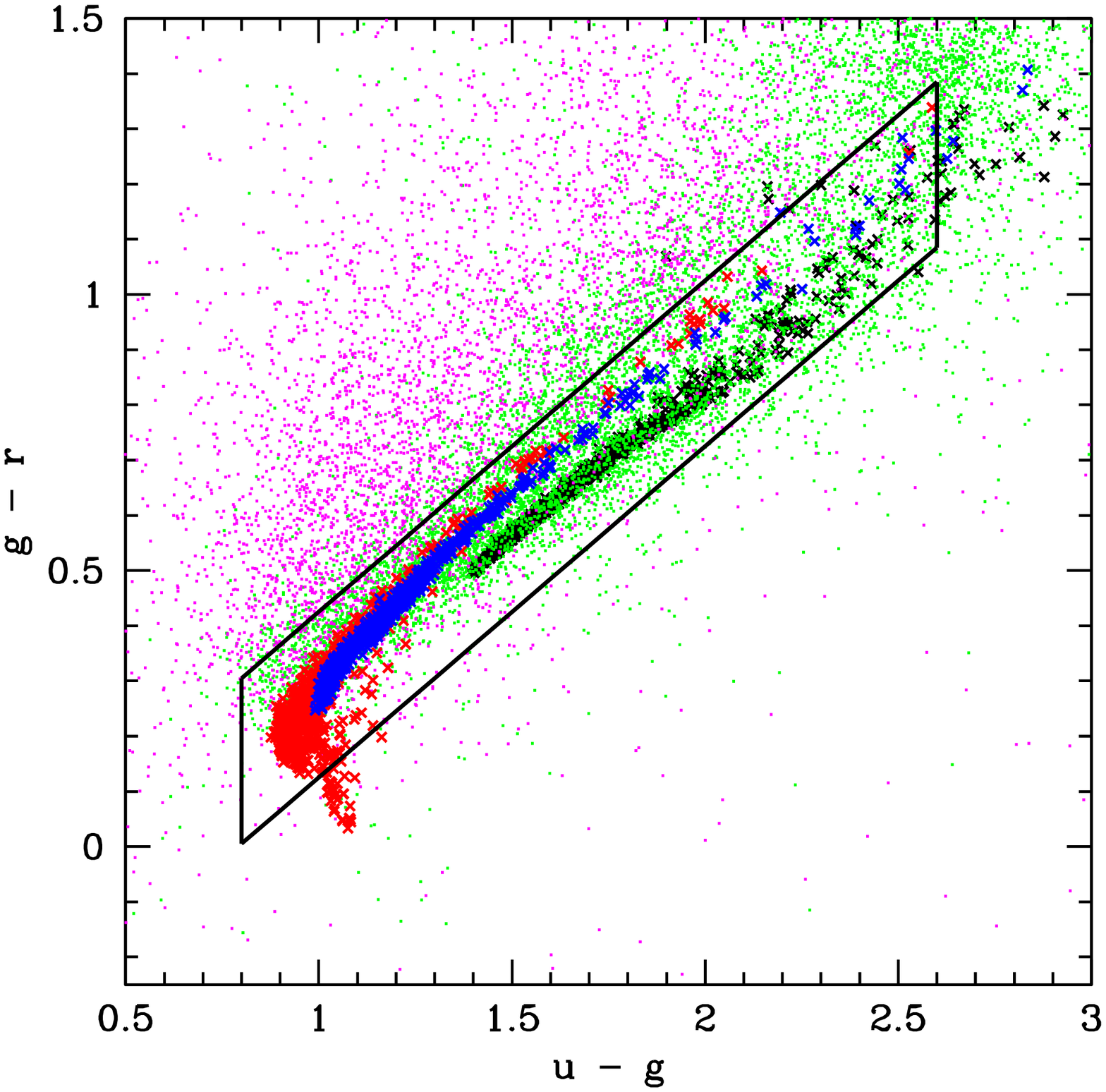}
  \centering \includegraphics[width=\columnwidth]{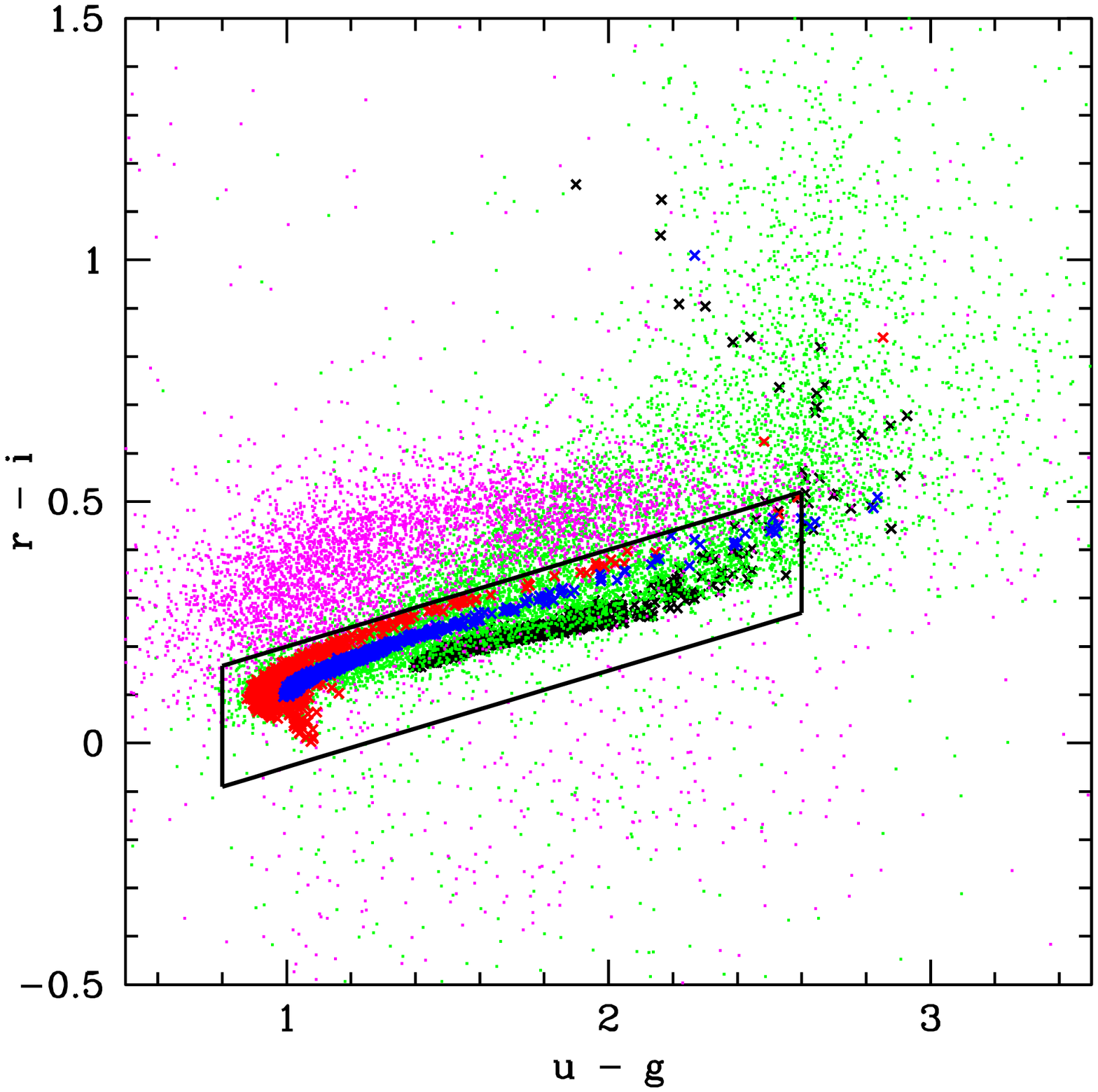}
  \caption{Color-color diagrams of model recoiled clusters.  Plotted
    is the distribution of recoiled star clusters with $N_{\rm cl} =
    100$ at $t_{\rm eject} \sim 10^{10}\,\yr$  with Z = $.02$ (Black Crosses), Z = $.0004$ (Red Crosses), and Varying Z (see text; Blue Crosses), along with a
    random selection of galaxies (magenta points) and stars (green
    points) from SDSS. The trapezoids correspond to our color-color
    selection (see \S~\ref{sec:photosearch}).}
  \label{fig:colorcolor}
\end{figure*}

In Figure~\ref{fig:colorcolor}, we plot a color-color diagram of our
Model Clusters with $N_{\rm cl} = 100$, and $t_{\rm eject} \sim
10^{10}\,\yr$ \newc{after ejection}.  Overall, the loci of star clusters follow the
distribution of galactic stars identified by SDSS, however, they
visually appear to be well separated from SDSS galaxies.  In the
figure, it is clear that many clusters contain a star on the giant
branch.  Because such stars are more luminous than all of the other
stars in the cluster, these systems are likely indistinguishable from the
population of late-type stars in the halo. Clusters with $N_{\rm cl} = 10 - 1000$ follow a similar distribution.

Because the star clusters can not be distinguished from individual
stars based on their colors alone, a successful photometric search
needs to exclude point sources, and use a magnitude system which
doesn't depend on the exact light profile of the object. We therefore
focus on the photometric properties of resolved objects, using the
Petrosian magnitude system \citep{2001AJ....121.2358B,
  2001AJ....122.1104Y}. This system is is defined by the Petrosian
ratio,
\begin{equation}
  \label{eq:petrosian}
  {\cal R}_P(r) \equiv \frac{\int_{0.8 r}^{1.25r} \rmd r' 2 \pi r' I(r')/[\pi (1.25^2-0.8^2)r^2]}{\int_0^r \rmd r' 2 \pi r' I(r')/(\pi r^2)},
\end{equation} 
where $I(r)$ is azimuthally averaged surface brightness profile in any
particular band. The Petrosian radius, $r_p$, is defined by ${\cal
  R}_p(r_p) = 0.2$ in the SDSS system. The total Petrosian flux
(and magnitude) from the object is calculated as the total integrated
light within $2 r_p$, where $r_p$ is determined in the $r$-band alone.

We use the azimuthally averaged cumulative light profile to
distinguish candidates from galaxies and point sources.  The SDSS
catalog has the mean flux of light in annuli around the peak of the
photometric object.  We add the light in these annuli to recreate the
total amount of light within radii $r_i \approx 0.22, 0.67, 1.03,
1.75, 2.97, 4.59, 7.36$ arcsec.  We use these bins to calculate the
logarithmic slope of the cumulative light profile, $\Gamma_i \equiv
\rmd \ln I_i / \rmd \ln r \approx \ln(I_{i+1}/I_i)/ \ln{(r_{i+1}/r_i)}
$, For a completely resolved star cluster with power-law density
profile $n \propto r^{-\alpha}$, we expect the light profile to
approach $3-\alpha \approx 0.85$ where $\alpha \approx 2.15$.  Outside
the half-light radius of the cluster, however, the slope of the
profile changes significantly from a single power-law.  To model this
we use the distribution of stars from our $N$-body simulations and
generate mock light density profiles with a variety of PSFs and
angular sizes.  To facilitate the search in \S~\ref{sec:photosearch},
we have split our mock profiles into bins of varying $r_p$.  The best
fit parameters are detailed in Table~\ref{tab:params}.

\begin{table}
 %\begin{minipage}[140mm]
  \caption{\label{tab:params}  Best fit SDSS light profiles to find mock clusters.  Because the cumulative light profiles are not corrected for seeing, we search for candidates by focusing on a range of properties that depend on the observed Petrosian Radius, $r_p$.  See \S~\ref{sec:models} for detailed definitions of the parameters.  ${^{\rm a}}$Seeing better than 1.2$\arcsec$. ${^{\rm b}}$Seeing between 1.2 and 1.7$\arcsec$. }
\footnotetext[2]{Seeing between 1.2 and 1.7}
\begin{tabular}{@{}lcccc@{}}
\hline
$r_{p}$  & $\Gamma_4$ & $\Gamma_5$ & $\Gamma_6$ & $\Gamma_7$\\
(arcsec) & &&&\\
\hline
2.0--3.0$^{\rm a}$         &     0.78 -- 0.88     &  ...   &   ... & ...    \\
2.0--3.2$^{\rm b}$          &     $\gtrsim .85 $    &  $\gtrsim 0.05$  &   ... &...     \\
3.0--4.5          &   ...          &    0.25 -- 0.5   &  0.10--0.25 &  ...   \\
4.0--6.0    &  ... &0.35 -- 0.70        &   0.28 -- 0.45          &  0.1 -- 0.2         \\

\end{tabular}
%\footnotetext[1]{Seeing better than 1.2}
%\footnotetext[2]{Seeing between 1.2 and 1.7}
%\end{minipage}
%\caption{$^{a}$ Seeing better than 1.2$\arcsec$}
\end{table}

\section{Photometric Search}
\label{sec:photosearch}
SDSS has imaged approximately one quarter of the sky to a limiting
magnitude $r \approx 22.2$. As the largest database of photometric and
spectroscopic objects in the sky , it presents a prime opportunity to
search for recoiled clusters.  In \S~\ref{sec:models} we developed a
simple photometric model of recoiled star clusters in Milky Way like
galaxies. Here we use this model to systematically search for
photometric candidates in SDSS DR7.

We use the SDSS DR7 {\small \sc CasJobs}\footnote{Located at
  \url{http://casjobs.sdss.org/CasJobs/}} photometry database to
select objects by size, shape, color, and azimuthally averaged light
profile. We limit our search to resolved objects with a
Petrosian radius $r_p>2\arcsec$ in the $g$ band.

Our color selection criteria is illustrated by the trapezoids in
Figure~\ref{fig:colorcolor}. We use the Petrosian magnitude system
color corrected for extinction \citep{1998ApJ...500..525S}. Our
criteria focus on an old population of metal-poor to solar metallicity
stars but excludes clusters with a star on the red giant branch. We
choose our candidate clusters out of the parallelogram defined by
$1.25 < u-g < 1.75$, $0.5 (u - g) - .225 < g - r < 0.5 (u-g) -0.075$.
Additionally, we require that the shape of the candidates be circular
by selecting for objects with ratio of semi-minor to semi-minor axes
greater than $0.7$.  These criteria select $\sim 70,000$ resolved
photometric objects as candidate recoiled clusters, with photometric
properties of stars.  We limit our sample further by using the
azimuthally averaged cumulative light profile ($\Gamma_i$) of the
remaining objects as detailed in Table~\ref{tab:params}. In addition
we search for simpler model clusters with $\Gamma_4$ and $\Gamma_5$
between 0.6 and 0.9.

\begin{figure*}
  \centering \mbox{\includegraphics[width=1.5in]{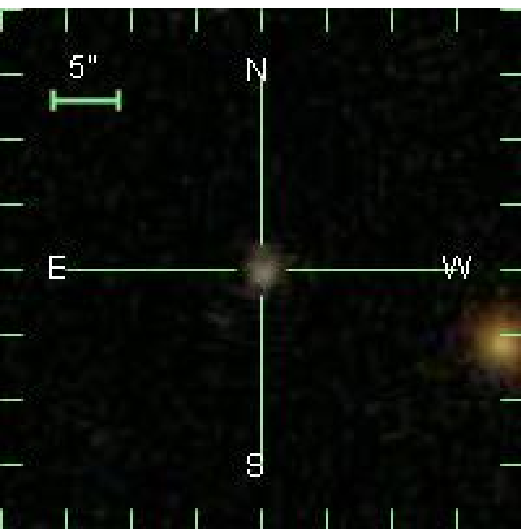}\hspace{10pt}
\includegraphics[width=1.5in]{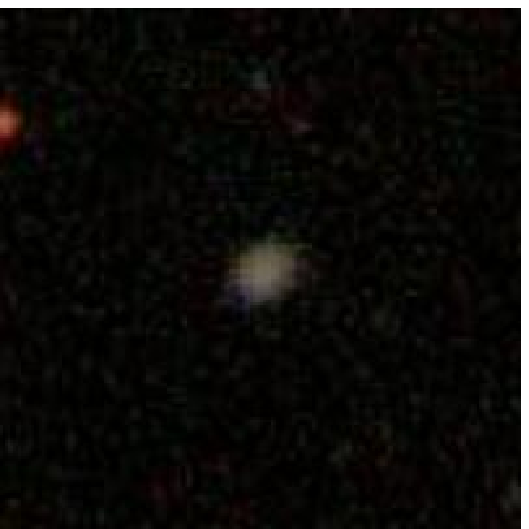}\hspace{10pt}
\includegraphics[width=1.5in]{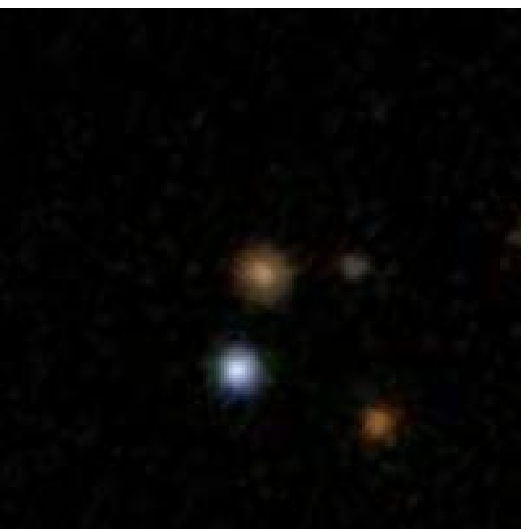}\hspace{10pt}
\includegraphics[width=1.5in]{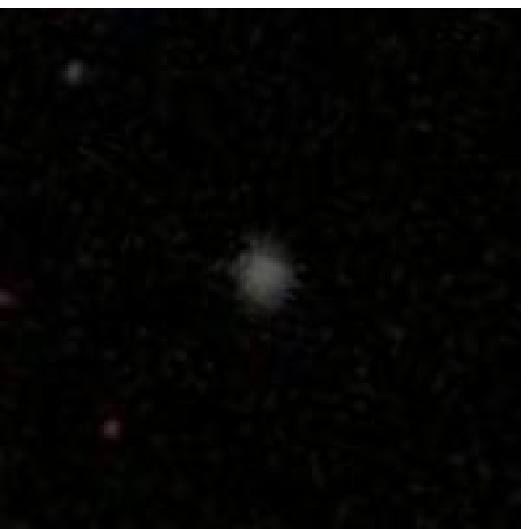}}
\mbox{\includegraphics[width=1.5in]{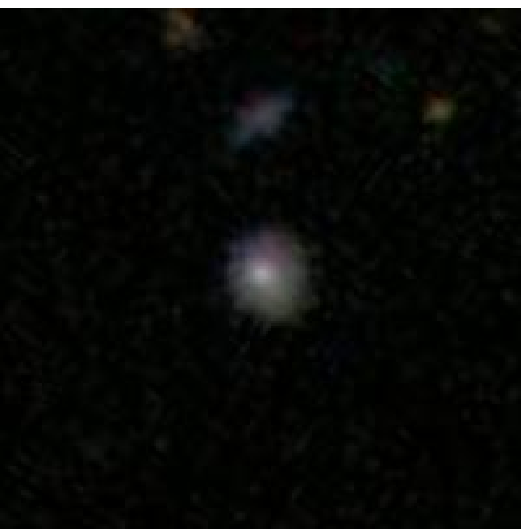}\hspace{10pt}
\includegraphics[width=1.5in]{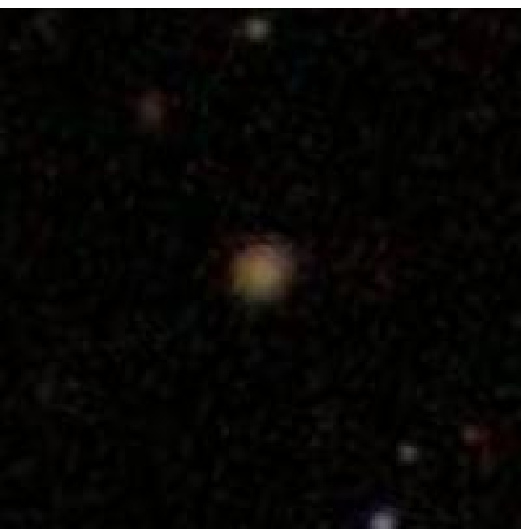}\hspace{10pt}
\includegraphics[width=1.5in]{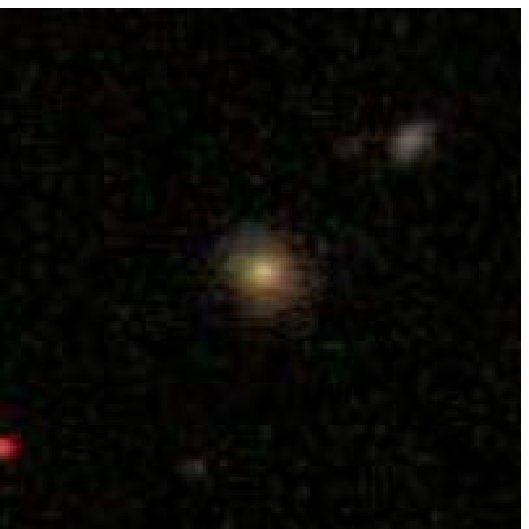}\hspace{10pt}
\includegraphics[width=1.5in]{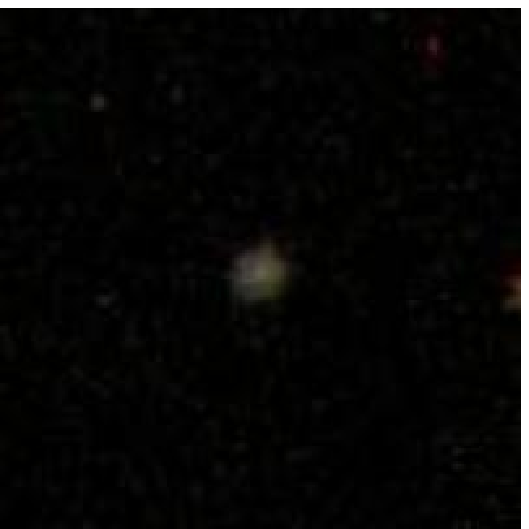}}
\caption{\label{fig:thumbs}Thumbnails of a diverse selection of candidates. From left to right and top to bottom these are 
SDSS J114607.52+135233.1,
SDSS J130154.22-031323.3,
SDSS J052222.65-013302.9,
SDSS J084034.69+162319.5,
SDSS J084822.47+355630.4,
SDSS J093815.82+231234.8,
SDSS J121414.73+161215.4,
and SDSS J123544.93+193016.9.  The scale is the same for all images, with the photometric object located at the center.
}
\end{figure*}

Using these criteria, we are left with $\sim 1,000$ candidates, which
we visually inspect to remove obvious interlopers.  These tend to be
individual or binary stars in crowded fields, face-on disk galaxies,
and cuspy elliptical galaxies.  We are left with $\sim 100$ objects, which we list in
Tables~\ref{tab:sdss} \& \ref{tab:slope}. Thumbnails of a selection of candidates is shown in Figure~\ref{fig:thumbs}. The number of candidates that
remain is likely a reflection of our ability to visually inspect the
candidates.  It is impossible to visually inspect the $70,000$
objects selected through color alone, and we had to use some model
dependent choices for the light profile to obtain a more reasonable
number of photometric candidates ($\sim 1,000$).

\begin{table}
\caption{\label{tab:sdss} Candidate recoil clusters based on the selection criteria described in \S~\ref{sec:photosearch} and Tab.~\ref{tab:params}}
\begin{tabular}{@{}lccc@{}}
\hline
Object & $g$ &$ r$ & $r_p$\\
\hline
SDSS J003550.53-100543.0  &  19.57  &  19.10  &  3.07\\
SDSS J005248.49+155331.6  &  19.81  &  18.99  &  3.24\\
SDSS J011023.54-090416.1  &  19.70  &  19.31  &  3.19\\
SDSS J015724.63-085424.1  &  18.48  &  18.19  &  3.30\\
SDSS J020705.55+003738.9  &  19.81  &  19.29  &  3.49\\
SDSS J021500.57+001217.8  &  20.40  &  19.77  &  3.57\\
SDSS J030347.44-081909.5  &  20.12  &  19.01  &  3.21\\
SDSS J064325.65+281559.3  &  20.51  &  19.80  &  2.79\\
SDSS J073940.32+221323.1  &  18.67  &  18.39  &  4.49\\
SDSS J074214.49+251424.0  &  20.14  &  19.35  &  2.94\\
SDSS J074827.56+261836.6  &  19.74  &  19.04  &  3.09\\
SDSS J075550.27+343959.3  &  20.51  &  19.89  &  3.09\\
SDSS J080005.57+514410.6  &  20.68  &  19.57  &  3.44\\
SDSS J081020.09+315018.6  &  20.72  &  19.93  &  2.75\\
SDSS J081546.83+155039.8  &  20.32  &  19.88  &  2.71\\
SDSS J082724.19+340543.7  &  19.77  &  19.23  &  3.43\\
SDSS J083158.70+332233.5  &  19.35  &  18.88  &  3.01\\
SDSS J083701.15+230023.2  &  20.54  &  20.20  &  2.56\\
SDSS J084034.69+162319.5  &  20.13  &  19.62  &  3.06\\
SDSS J084246.26+361533.7  &  19.66  &  19.09  &  4.00\\
SDSS J084505.56+451932.0  &  20.59  &  20.11  &  3.35\\
SDSS J084647.94+001638.4  &  21.07  &  20.44  &  3.43\\
SDSS J092335.02+472837.1  &  19.64  &  19.01  &  2.99\\
SDSS J092607.17+021555.4  &  20.39  &  19.60  &  3.28\\
SDSS J092757.48+054543.7  &  20.03  &  19.22  &  3.12\\
SDSS J092921.02+545144.3  &  19.31  &  18.97  &  3.25\\
SDSS J093815.82+231234.8  &  20.10  &  19.19  &  3.06\\
SDSS J094801.22+324203.4  &  19.72  &  19.13  &  3.02\\
SDSS J101754.64+803827.9  &  20.02  &  19.14  &  3.59\\
SDSS J104012.45+183600.5  &  19.71  &  19.17  &  3.12\\
SDSS J104012.45+645611.0  &  20.20  &  19.45  &  3.13\\
SDSS J104700.20+451459.4  &  19.95  &  19.24  &  3.31\\
SDSS J105840.27-012816.9  &  20.53  &  19.71  &  2.99\\
SDSS J105846.93+170430.1  &  19.07  &  18.37  &  3.19\\
SDSS J105907.75-031445.6  &  19.60  &  18.67  &  3.48\\
SDSS J111223.68+072718.0  &  19.23  &  18.70  &  3.87\\
SDSS J111327.28+081717.1  &  18.88  &  18.55  &  4.30\\
SDSS J112142.07+182723.3  &  19.53  &  19.12  &  3.23\\
SDSS J112711.19+113814.8  &  20.14  &  19.24  &  3.54\\
SDSS J113013.20+643939.4  &  18.99  &  18.72  &  3.76\\
SDSS J113137.97+172219.0  &  19.49  &  18.66  &  3.49\\
SDSS J113308.62+002113.2  &  20.84  &  19.71  &  2.82\\
SDSS J114546.59+081137.7  &  19.67  &  19.08  &  3.87\\
SDSS J115108.59+030704.8  &  19.63  &  19.33  &  2.92\\
SDSS J115253.98+171842.6  &  20.00  &  19.27  &  2.92\\
SDSS J115526.94+355320.0  &  19.95  &  19.55  &  3.01\\
SDSS J115543.73+333639.9  &  18.83  &  18.39  &  4.43\\
SDSS J115957.30+020749.5  &  19.72  &  19.14  &  3.27\\
SDSS J120446.11+270030.1  &  19.94  &  19.42  &  3.06\\
SDSS J120533.94+022352.9  &  20.13  &  19.28  &  3.11\\

\end{tabular}
\end{table}

\begin{table}
\addtocounter{table}{-1}
\caption{ \bf continued.}

\begin{tabular}{@{}lccc@{}}
\hline
Object &$ g $& $r$ & $r_p$\\
\hline
SDSS J120648.21+450646.7  &  21.06  &  20.08  &  3.16\\
SDSS J121700.34+353542.1  &  20.38  &  19.74  &  3.22\\
SDSS J123544.93+193016.9  &  20.55  &  19.87  &  2.84\\
SDSS J123614.93+013708.7  &  20.27  &  19.81  &  3.03\\
SDSS J125011.62-021800.1  &  21.16  &  20.51  &  2.65\\
SDSS J125734.92+253916.5  &  20.49  &  19.92  &  2.78\\
SDSS J125915.28+070342.6  &  20.01  &  19.52  &  3.30\\
SDSS J125958.63-002508.4  &  19.75  &  18.71  &  3.26\\
SDSS J130109.32+462607.1  &  21.28  &  20.15  &  2.76\\
SDSS J130153.77+504842.2  &  20.38  &  19.64  &  2.85\\
SDSS J130154.22-031323.3  &  19.67  &  18.97  &  3.32\\
SDSS J132249.56+084115.6  &  19.85  &  19.22  &  3.27\\
SDSS J132610.71+535511.7  &  20.79  &  20.26  &  3.24\\
SDSS J133057.40+184836.7  &  20.00  &  19.28  &  2.66\\
SDSS J133212.59+353159.7  &  19.56  &  18.94  &  2.93\\
SDSS J134127.13+081550.6  &  21.56  &  20.52  &  2.06\\
SDSS J134459.96+030428.6  &  19.76  &  19.55  &  3.13\\
SDSS J134737.53+203427.0  &  20.19  &  19.78  &  2.94\\
SDSS J134852.16+245743.4  &  19.72  &  19.16  &  3.77\\
SDSS J135040.46+103538.7  &  20.75  &  19.95  &  2.84\\
SDSS J135241.53+121430.8  &  20.51  &  19.66  &  3.06\\
SDSS J135544.47-065531.4  &  20.29  &  19.54  &  2.90\\
SDSS J140113.91+060627.7  &  20.13  &  19.71  &  2.85\\
SDSS J141327.28+282847.1  &  19.58  &  18.79  &  3.05\\
SDSS J141418.26+454312.8  &  18.94  &  18.57  &  4.37\\
SDSS J142920.56+261616.5  &  20.13  &  19.39  &  2.80\\
SDSS J142935.43+073722.6  &  19.59  &  18.87  &  3.18\\
SDSS J145030.79+380441.6  &  19.62  &  19.06  &  2.91\\
SDSS J145145.53+103402.0  &  19.94  &  19.32  &  3.45\\
SDSS J145150.02+352929.8  &  20.03  &  19.53  &  2.78\\
SDSS J145345.50+080808.7  &  20.37  &  19.58  &  3.03\\
SDSS J150113.32+051304.1  &  19.92  &  19.37  &  2.91\\
SDSS J150459.72+081819.7  &  20.50  &  19.89  &  2.78\\
SDSS J151934.33+134102.8  &  20.51  &  20.12  &  3.33\\
SDSS J152006.92+085031.0  &  20.45  &  19.68  &  2.78\\
SDSS J152249.28+473700.1  &  20.31  &  19.82  &  3.12\\
SDSS J152646.00+210607.1  &  21.21  &  19.97  &  3.33\\
SDSS J153145.65+150057.6  &  20.12  &  19.54  &  3.11\\
SDSS J155333.07+423146.0  &  20.25  &  19.53  &  3.07\\
SDSS J155442.55+055111.1  &  19.09  &  18.60  &  3.23\\
SDSS J160630.13+351046.2  &  19.48  &  19.06  &  3.04\\
SDSS J160702.48+110353.3  &  20.27  &  19.78  &  3.89\\
SDSS J162536.57+563531.9  &  20.99  &  19.96  &  3.53\\
SDSS J163339.07+132635.6  &  20.23  &  19.56  &  3.46\\
SDSS J163659.29+235816.2  &  20.05  &  19.01  &  2.93\\
SDSS J170525.39+235241.5  &  19.17  &  18.37  &  4.08\\
SDSS J172243.89+080447.8  &  20.44  &  19.97  &  2.91\\
SDSS J210803.13-001350.4  &  20.35  &  19.52  &  3.15\\
SDSS J213035.54-070545.7  &  20.76  &  20.05  &  3.19\\
SDSS J215424.98+002023.4  &  20.01  &  19.14  &  3.23\\
SDSS J233106.11+075810.9  &  20.89  &  19.99  &  2.97\\

\end{tabular}
\end{table}

\begin{table}
\caption{\label{tab:slope}  Candidate recoil clusters based on an asymptotic cumulative light profile with slope between 0.6 and 0.9.}

\begin{tabular}{@{}lccc@{}}
\hline
Object & $g $& $r$ & $r_p$\\
\hline
SDSS J084822.47+355630.4  &  18.69  &  18.21  &  3.60\\
SDSS J045505.26+244156.3  &  21.06  &  19.43  &  2.94\\
SDSS J114607.52+135233.1  &  20.57  &  19.81  &  2.53\\
SDSS J090546.22+225309.9  &  20.50  &  19.77  &  2.88\\
SDSS J102509.23+215445.8  &  20.22  &  18.86  &  3.84\\
SDSS J211006.40+002759.2  &  19.18  &  18.60  &  3.32\\
SDSS J100310.59+282625.0  &  18.86  &  18.05  &  3.45\\
SDSS J121414.73+161215.4  &  19.15  &  18.22  &  3.63\\
SDSS J051413.98+164920.2  &  18.95  &  17.89  &  3.41\\
SDSS J161526.75+110822.9  &  21.62  &  20.84  &  2.37\\
SDSS J052222.65-013302.9  &  20.16  &  19.17  &  2.77\\
SDSS J135018.11+092421.7  &  18.89  &  17.90  &  3.79\\
SDSS J003209.31+071259.2  &  18.94  &  17.80  &  5.22\\
SDSS J144115.60+185843.8  &  20.12  &  19.26  &  3.46\\
SDSS J160236.41+322318.7  &  19.88  &  18.75  &  2.87\\
SDSS J163339.50+440918.3  &  20.02  &  18.98  &  2.57\\
SDSS J132908.57+232303.8  &  19.62  &  18.46  &  3.09\\
SDSS J144856.99+153744.6  &  20.39  &  19.54  &  2.81\\

\end{tabular}
\end{table}

\section{Spectroscopic Search}
\label{sec:specsearch}
The selection of spectroscopic objects in SDSS DR7 is not ideal for
serendipitously locating the star cluster around a recoiled BH.
Indeed, many of the main science objectives for spectroscopic targets
specifically exclude objects with the photometric properties similar
to the recoiled clusters we search for. Nevertheless, resolved
recoiled clusters would be identified in SDSS photometrically as a
galaxy because of its extended size, but unlike galaxies, would have
an extremely low redshift, $z \lesssim 10^{-3}$.

We use SDSS spectroscopy to select candidate clusters by their
identified redshift with $-0.002 < z < 0.002$, corresponding to radial
velocities with magnitude less than $600\,\kms$. To exclude single and
unresolved binary stars, we restrict our results to objects with
Petrosian radii, $r_p > 3.0\,\arcsec$ in both the $r$ and $g$
bands. We then remove blended objects, which are mostly galaxies with
foreground stars. This results in approximately $270$ objects
identified by SDSS photometry as galaxies, and $18$ identified as
stars.  We have visually inspected all $290$ objects. All of the
remaining objects in the sample have two main features: (i) a star
like object on a diffuse source or (ii) featureless and diffuse
source.  In case (i), the spectral identification of the source is
always stellar.  The majority of objects that fall into category (ii)
were incorrectly identified with $z=0$.  The remaining cases were not
spherical in shape or clumpy, and therefore not classified as recoiled
cluster candidates.

\section{Literature Search}
\label{sec:histsearch}

We have also explored the literature for galaxies that were
spectroscopically identified as stars owing to their low
redshift. \citet{1970ApJ...160..405S} took low-dispersion spectra
of 141 objects selected from Zwicky's catalogs of compact galaxies
\citep[available in][]{1971cscg.book.....Z}, and found that 14 objects
had near-zero redshifts and identified the galaxies as having a
foreground star. We have reanalyzed newer digital images and spectra
of these objects. In some cases the foreground star has moved and new
spectra show the objects are extragalactic.  However, most were
visually identified as galaxies because of their disk like shape. Only one
object, IV Zw 26, could not be excluded as a candidate owing to
insufficient resolution in any survey.  

\section{Summary and Conclusions}
\label{sec:conclusions}

We have followed the long term evolution of star clusters around
recoiled BHs using long term $N$-body simulations, with a one-to-one
correspondence between the stars and $N$-body particles. We have found
that for $\msmbh = 10^4\,\msun$, $\sim 40\%$ of the stars are ejected
from the star cluster, and $\sim 40\%$ of the stars are tidally
disrupted by the central BH within $10^{10}\,\yr$.  We have scaled
these results to BHs with masses $\msmbh \lesssim 2\times 10^6\,\msun$,
finding that $N_{\rm cl} = 840 (\msmbh / 10^5\,\msun)^{13/8}$ stars
remain around the BH today for a typical recoiled BH.  For more
massive BHs, the cluster should eject \newc{or disrupt} few stars \newc{over $\sim 10^{10}\,yr$.} Although a single BH has a
small tidal disruption rate, we have found that the total rate for all
clusters in Milky Way like galaxies is $\sim 10^{-7}\,\yr^{-1}$, which
is only a factor of $\sim 10$ lower than \newc{expected} in the galactic
center. \newc{We have extended our one-dimensional Fokker-Planck treatment in Paper I
  to include resonant relaxation and large-angle scattering to account
  for the dominate mass  loss mechanisms of the cluster.  Using this
  treatment, we were able to get satisfactory agreement between the
  $N$-body simulations and Fokker-Planck simulations. Some discrepancy
remained, which we attribute to the large amount of anisotropy in
realistic clusters.}

We used our $N$-body simulations to generate random realizations of
star clusters today, which guided our search for star clusters around
recoiled BHs in SDSS.  In our photometric search through SDSS data, we
assumed that the star clusters have a power-law density profile and
that they have colors comparable to an old population of stars.  We
used these criteria to find $\sim 70,000$ candidates of which only
$\sim 1000$ had a light density profile out to $4\arcsec$ consistent
with a recoiled star cluster with power-law density slope. We visually
inspected all candidates, and found that many were the bulges of
nearby face-on spirals.  The remaining $100$ candidates were faint,
and difficult to distinguish from distant galaxies.  Follow-up
spectroscopy is necessary to identify their nature. If any of them are
a star cluster around a recoiled BH, it would show unusually high
velocity dispersion $\sigma_\star$ at $z \approx 0$, with a spectrum
of a population of old stars.

We also searched the spectroscopic database of SDSS for resolved
objects with a low redshift/blueshift consistent with the 
Local Group.  The vast majority of these candidates were galaxies with
bright foreground stars.  We found no candidates that appeared to be a
recoiled cluster.

The criteria we used to search for candidate clusters required the
cluster to have a well defined power-law density profile, as found
in our numerical simulations.  Because we did not include compact
remnants in our simulations (which would have the correct density
profile but not light profile) we can not be confident that we
properly included all star clusters in our search.
Unfortunately, most research on stellar remnants around supermassive
black holes focus on mass segregation around relaxed stellar cusps
\citep{2006ApJ...649...91F,2006ApJ...645L.133H,2009MNRAS.395.2127O,2009ApJ...697.1861A,2009ApJ...698L..64K}.
These simulations find that compact remnants play an important role in
the dynamics for radii $r \lesssim r_k$.  We have not included these
objects because it is difficult to quantify how many compact remnants
would be in this region immediately before the binary merges.  Indeed,
the segregation of the compact remnants can only occur after the
low-mass stars populate this region \citep{2009arXiv0909.1318M}.  In
future work we hope to include these compact remnants, and extend our
Fokker-Planck code to include large angle scattering between stars of
multiple masses \citep{2009MNRAS.395.2127O}.

 An alternative search strategy, which we did not explore here, is to
 cross-check less stringent criteria with alternative databases, such
 as the ROSAT {\em x-ray} survey.

\section*{Acknowledgments}
We thank Warren Brown, Kelly Holley-Bockelmann, Nick Stone, and Matt Walker for
their helpful discussions. We thank Wallace Sargent and Francois
Schweizer for bringing Zwicky's catalog to our attention. We thank Ulf
L{\"o}ckmann for publicly releasing his {\small \sc BHint} code, as
well as the {\small \sc BaSTI} group for allowing us to use their
code.  This project makes use of data products from the Sloan Digital
Sky Survey, which is managed by the Astrophysical Research Consortium
for the Participating Institutions.  This research has also made use
of the NASA/IPAC Extragalactic Database (NED) which is operated by the
Jet Propulsion Laboratory, California Institute of Technology, under
contract with the National Aeronautics and Space Administration.
This work was supported in part by NSF grant AST-0907890 and NASA
grants NNX08AL43G and NNA09DB30A.  Support for the completion
of this work was provided to RMO by the National Aeronautics and Space
Administration through Einstein Postdoctoral Fellowship Award Number
PF0-110078 issued by the Chandra X-ray Observatory Center, which is
operated by the Smithsonian Astrophysical Observatory for and on
behalf of the National Aeronautics Space Administration under contract
NAS8-03060

\bibliography{p}

\end{document}